\documentclass[%
 aip,
 amsmath,amssymb,
 reprint,%
]{revtex4-1}

\usepackage{graphicx}
\usepackage{dcolumn}
\usepackage{bm}

\usepackage[utf8]{inputenc}
\usepackage[T1]{fontenc}
\usepackage{etoolbox}
\usepackage{amsmath}
\usepackage{amssymb}

\usepackage{float}
\usepackage[noend]{algpseudocode}

\usepackage{algorithmicx,algorithm}

\makeatletter
\def\@email#1#2{%
 \endgroup
 \patchcmd{\titleblock@produce}
  {\frontmatter@RRAPformat}
  {\frontmatter@RRAPformat{\produce@RRAP{*#1\href{mailto:#2}{#2}}}\frontmatter@RRAPformat}
  {}{}
}%
\makeatother
\begin{document}

\preprint{AIP/123-QED}

\title[Published at the Physics of Fluids]{How to Control Hydrodynamic Force on Fluidic Pinball via Deep Reinforcement Learning}

\author{Haodong Feng}
\thanks{Equally contribution.}
\affiliation{ 
Zhejiang University, Hangzhou, Zhejiang 310027, China.
}
\affiliation{School of Engineering, Westlake University, Hangzhou, Zhejiang 310030, China.}
\affiliation{ 
Microsoft Research AI4Science, Beijing 100080, China.
}

\author{Yue Wang}
\thanks{Equally contribution.}
\affiliation{ 
Microsoft Research AI4Science, Beijing 100080, China.
}

\author{Hui Xiang}
\affiliation{Scien42.tech, Beijing 100101, China.}

\author{Zhiyang Jin}
\thanks{The authors to whom correspondence may be addressed: fandixia@westlake.edu.cn, jinzhiyang94@163.com.}
\affiliation{Mechanical and Electrical Engineering College, Hainan University, Haikou, Hainan 570228, China.}
\affiliation{Hainan Policy and Industrial Research Institute of Low-Carbon Economy, Haikou, Hainan 570228, China.}

\author{Dixia Fan}
\thanks{The authors to whom correspondence may be addressed: fandixia@westlake.edu.cn, jinzhiyang94@163.com.}
\affiliation{Research Center for Industries of the Future, Westlake University, Hangzhou, Zhejiang 310030, China.}
\affiliation{Key Laboratory of Coastal Environment and Resources of Zhejiang Province, School of Engineering, Westlake University, Hangzhou, Zhejiang 310030, China.}
\affiliation{School of Engineering, Westlake University, Hangzhou, Zhejiang 310030, China.}



\begin{abstract}
Deep reinforcement learning (DRL) for \textit{fluidic pinball}, three individually rotating cylinders in the uniform flow arranged in an equilaterally triangular configuration, can learn the efficient flow control strategies due to the validity of self-learning and data-driven state estimation for complex fluid dynamic problems. In this work, we present a DRL-based real-time feedback strategy to control the hydrodynamic force on \textit{fluidic pinball}, i.e., force extremum and tracking, from cylinders' rotation. By adequately designing reward functions and encoding historical observations, and after automatic learning of thousands of iterations, the DRL-based control was shown to make reasonable and valid control decisions in nonparametric control parameter space, which is comparable to and even better than the optimal policy found through lengthy brute-force searching. Subsequently, one of these results was analyzed by a machine learning model that enabled us to shed light on the basis of decision-making and physical mechanisms of the force tracking process. The finding from this work can control hydrodynamic force on the operation of \textit{fluidic pinball} system and potentially pave the way for exploring efficient active flow control strategies in other complex fluid dynamic problems.

\end{abstract}

\maketitle

\section{\label{sec:level1}Introduction}

Control of fluid flows has been a long-standing challenge for the engineering community \cite{kerswell2018nonlinear, rowley2017model, ashill2005review, choi2008control}. With the novel applications appearing in environment and energy engineering, its importance has increased. Among these challenges, active flow control (AFC) has been an attractive topic in fluid dynamics, where actuators intentionally change the fluid system by applying optimized inputs \cite{xu2023reinforcement, greenblatt2022flow, jovanovic2021bypass, yeh2021network, greco2020karman, gutmark1999flow, kim2007linear, cattafesta2011actuators, shaqarin2018need, kaul2022active}. Compared with the passive flow control (PFC) strategies usually involving structural optimizations, AFC is adaptive and can achieve more effective control in a broader working range. If the information from the system output is fed back to guide the control, AFC can be divided into open-loop control or closed-loop control. Unlike the open-loop control, the closed-loop control can utilize the feedback signals with the current features, to adjust the actuator in real-time, thus allowing automatic operation in a broader flow range and higher accuracy \cite{linkmann2020linear, aamo2003flow}.

In recent years, AFC research has started to pay close attention to the \textit{fluidic pinball} problem \cite{ishar2019metric, deng2020low} (shown in FIG. 1) with a simple configuration but rich physics, making it a benchmark testbed for AFC algorithms \cite{li2022machine, deng2018route, pastur2018reduced}. With different rotation velocities, this simple arrangement demonstrates a versatile set of flow patterns, categorized into six wake stabilization strategies: phasor control, boat-tailing, base-bleed, Magnus effect, especially high-frequency forcing, low-frequency forcing \cite{cornejo2019artificial}. The domestication of \textit{fluidic pinball} must consider the richness of these driving mechanisms. In addition, it allows the testing of steady and unsteady control strategies in a larger control parameter space. Therefore, the \textit{fluidic pinball} is a heuristic experiment suitable for developing advanced control methods to explore large state and action spaces for various objectives.


Various researchers have applied different AFC algorithms on the \textit{fluidic pinball}, illustrating the effectiveness of the problem setup and the corresponding control strategies. Peitz \textit{et al.} \cite{peitz2020data} developed the model predictive control (MPC) approach via the Koopman operator to control the lift of all three cylinders by rotating the top and bottom cylinders. Raibaudo \textit{et al.} \cite{raibaudo2020machine, raibaudo2021unsteady} applied the linear genetic algorithm (GA) to reduce the drag and modify the wake of \textit{fluidic pinball}. Blanchard \textit{et al.} \cite{blanchard2021bayesian} presented the performance of Bayesian optimization (BO) to design an open-loop controller for drag reduction. Another open-loop controller was developed by Ghraieb \textit{et al.} \cite{ghraieb2021single} applying the single-step deep reinforcement learning (DRL) as an optimization method. Maceda \textit{et al.} \cite{maceda2021stabilization} proposed a gradient-enriched machine learning control (gMLC) to stabilize the unstable symmetric wake behind the \textit{fluidic pinball}, which can explore new minima while keeping the convergence efficiency combining the merits of exploitation and exploration. Furthermore, the explorative gradient method (EGM) was proposed by Li \textit{et al.} \cite{li2022explorative} to avoid suboptimal local minima when minimizing the drag of \textit{fluidic pinball}. These control strategies can be employed, not only to optimize large parameter space, but also to reveal unknown solutions or parameter relationships for AFC in \textit{fluidic pinball}.

\begin{figure*}
  \centerline{\includegraphics[width=1\textwidth]{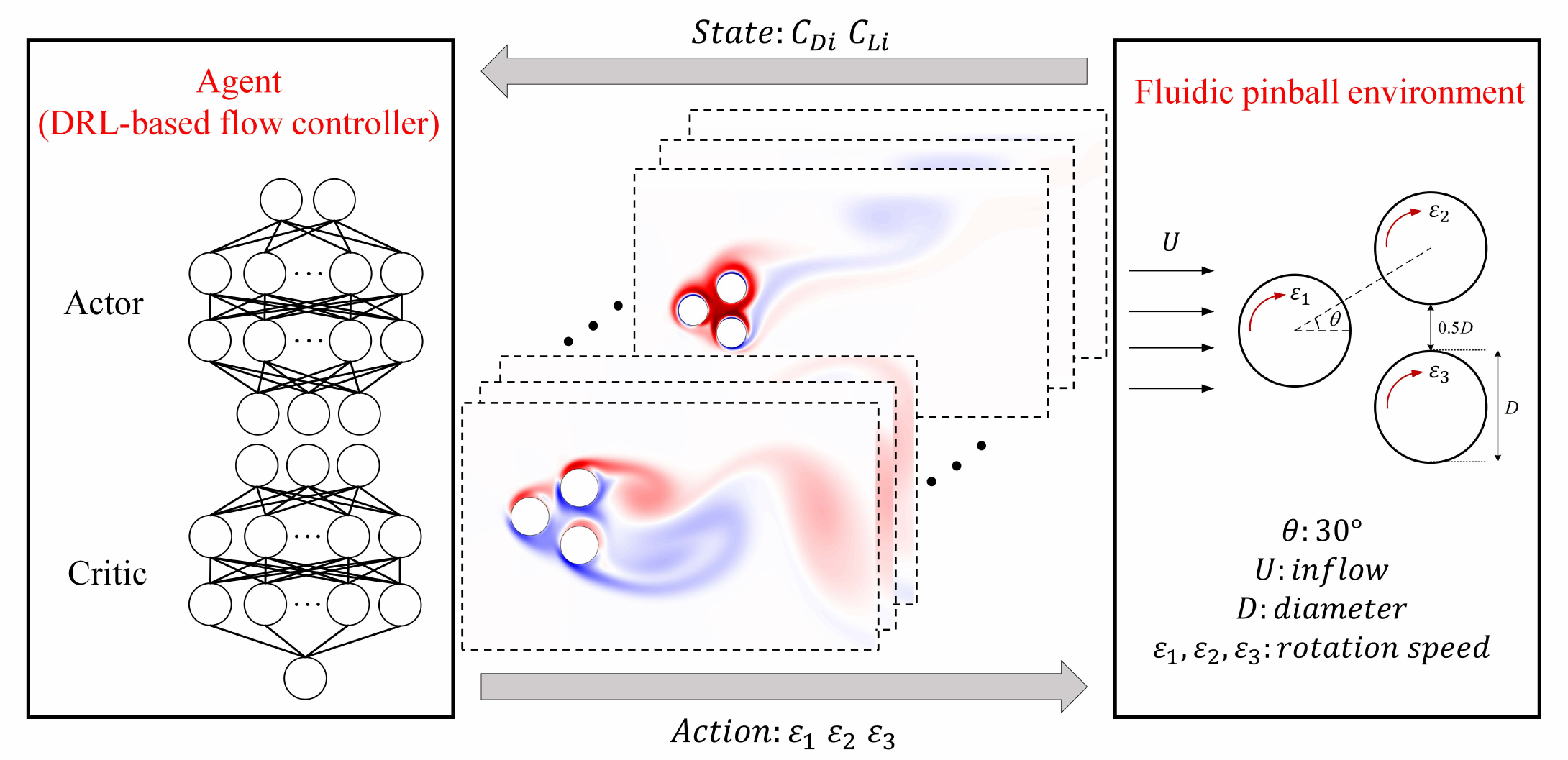}}
  \caption{\textbf{Framework of the DRL-based flow control for the \textit{fluidic pinball} environments.} The neural network in lift is the DRL agent and the system in right is the environment. The vorticities figures located at center intuitively demonstrate that flow can be controlled in real-time, to achieve the expected objective. In each episode, the agent interacts with the environment via the XML-RPC protocol at a fixed frequency. Each interaction consists of the state $C_{Di}$ \& $C_{Li}$ feedback from the environment that is determined by the objective of tasks, and action $\varepsilon_{i}$, $i = 1,2,3$. After collecting the data from the environment, the agent updates the policy while waiting for the reset of the environment. Along with control, a significant change in vortical wake occurs, shown in the snapshots boxed in the dashed line.}
\label{fig1}
\end{figure*}

Apart from work mentioned above, the rapid development of deep learning (DL) methods has paved the way for high dimensional and complex fluid problems \cite{brenner2019perspective,brunton2020machine,brunton2022applying}. Lee \textit{et al.} \cite{lee2019data} applied DL networks to predict flow field development for the canonical unsteady vortex shedding over a circular cylinder. Kharazmi \textit{et al.} \cite{kharazmi2021inferring} developed robust physics-informed neural networks (PINNs) to forecast the motion of a flexible cylinder undergoing vortex-induced vibrations (VIVs). Furthermore, the development of DL also got a foundation and powerful impetus for DRL \cite{sutton2018reinforcement}, which simplifies continuous systems into Markov decision processes (MDP) \cite{sutton1999between}. The DRL algorithm has been applied in various domains, including robotics, computer games, and autonomous driving \cite{silver2018general,aradi2020survey, wang2020irda, singh2022reinforcement}. DRL algorithm explores the control strategy through the interaction with the environment shown in FIG. \ref{fig1}, which is quite convenient and suitable for complex fluid dynamics. In recent years, due to the enormous potential of intensive DRL algorithms, which can deal with strongly nonlinear multi-scale dynamics problems \cite{feng2021optimal}, many pieces of research have been published on using DRL algorithms to solve AFC problems \cite{chen2022review}.

Many DRL applications for AFC can be found in several reviews by Brunton \textit{et al.} \cite{brunton2020machine}, Rabault \textit{et al.} \cite{rabault2020deep}, Viquerat \textit{et al.} \cite{viquerat2022review} and Chen \textit{et al.} \cite{chen2022review}. In particular, A closed-loop control method for synthetic jets was presented by Rabault \textit{et al.} \cite{rabault2019artificial} to reduce the drag of a cylinder at different Reynolds numbers. Xu \textit{et al.} \cite{xu2020active} used a feasible DRL [proximal policy
optimization, (PPO)] algorithm to look for the wake stabilization mechanism of a 2-D circular cylinder by tuning the rotations of two small cylinders behind it. A similar setup in the experiment was conducted by Fan \textit{et al.} \cite{fan2020reinforcement}. Zheng \textit{et al.} \cite{zheng2021active} adopted both active learning and DRL based on two jet actuators to eliminate vortex shedding and VIVs at low Reynolds numbers. Wang \textit{et al.} \cite{wang2022drlinfluids} proposed a python platform for coupling DRL and OpenFOAM in fluid mechanics, then demonstrated its reliability and efficiency using two wake stabilization benchmark problems. Han \textit{et al.}, Amico \textit{et al.}, Mei \textit{et al.} and Wang \textit{et al.} \cite{han2022deep, amico2022deep, mei2022active, wang2022accelerating} used DRL to manipulate the wake and hence reduce the drag of bluff bodies. Using expert demonstration from a wake oscillator, Zheng \textit{et al.} \cite{zheng2022data} demonstrated VIV reduction using DRL. Ren \textit{et al.} \cite{ren2021applying}, Ren and Tang \cite{ren2021bluff} developed the DRL-based AFC methods for bluff bodies to hide their hydrodynamic traces using a group of windward-suction-leeward-blowing (WSLB) actuators. Xie \textit{et al.} \cite{xie2022active} employed the DRL method to control a heaving plate breakwater to find the optimal wave dissipation policy. Qin \textit{et al.} introduced the dynamic mode decomposition (DMD) in reward function construction for DRL, so that the agent can learn the AFC policy through the more global information of the field. Li and Zhang \cite{li2022reinforcement} studied the application of a PPO control algorithm to the flow past a cylinder between two walls in order to suppress vortex shedding successfully. Pino \textit{et al.} \cite{pino2022comparative} systematically compared and analyzed the algorithms of machine learning and DRL on three typical AFC problems. Paris \textit{et al.} \cite{paris2021robust}, Xu and Zhang \cite{xu2023reinforcement} also discussed the optimal sensor placement along with the active flow control tasks with DRL. There are also many effective and successful cases of active flow control via DRL, considering various Reynolds numbers and tasks \cite{varela2022deep, guastoni2023deep}. The above successful applications of DRL in AFC further make us believe DRL algorithms can solve some control problems in fluid dynamics efficiently with the advantages of self-learning, data-driven, dimensional mapping, and generalization capability. They have shown great potential achieving force control for more complex and more challenging fluidic problems.

At the time of submission, most DRL applications were bound to search and solve for lower-dimension and less-mode flow control problems. In this work, we applied the DRL-based control method on a more complex fluid benchmark problem (\textit{fluidic pinball}), with a bigger nonparametric action space, demonstrating its feasibility and future challenges. We present applying DRL-based real-time feedback control to minimize the drag, maximize the lift, and even remain the drag to the expected references in real-time for the \textit{fluidic pinball}. Although some articles have published research on drag reduction, the force tracking as expected freely for \textit{fluidic pinball} still has an exciting gap waiting to be explored, which is focused by seldom previous work. Moreover, this work is also the first attempt to apply the DRL algorithm for the real-time feedback force control on \textit{fluidic pinball} setup for extremum searching and force tracking. The results illustrate that the DRL algorithm is suitable and feasible for challenging multi-input multi-output (MIMO) control problems like the present work.

The paper is organized as follows. In section II, the problem setup, the computational fluid dynamics (CFD) solver methodologies, and the open-loop and DRL-based real-time control approaches are described. An introduction to the DRL algorithm is also presented. In section III, the results of open-loop control with constant rotation speed from brute-force searching are analyzed, focusing firstly on the introduction of baselines that will compare with the results of DRL-based control. Secondly, a detailed discussion of force distribution and vortical wake of \textit{fluidic pinball} with overall discrete control rotations is reported. In section IV, the results of drag reduction and drag tracking problems solved via DRL-based real-time control are presented, and the addition comments of drag tracking are discussed. The conclusion follows in section V.

\section{\label{sec:level1}Materials and Methods}

\subsection{\label{sec:level2}Problem Setup}

In this work, we focus on the \textit{fluidic pinball} problem of a 2-D flow past three circular cylinders of diameter $D$, arranged in an equilaterally triangular configuration \cite{maceda2022xmlc} at Reynolds number $Re = UD/\nu = 100$, where $U$ is the incoming velocity, and $\nu$ is the kinematic viscosity of the fluid. A sketch of the problem setup is shown in FIG. \ref{fig1}. Specifically, The incoming flow from the inlet boundary is uniform. All results will be nondimensionalized by $U$, $D$, and time $T$. Determined by the DRL agent in real-time, the three cylinders can rotate at a speed of $\varepsilon_i \in [-5, 5]$, where $\varepsilon_i = \frac{\omega D}{U}$, $\omega$ is the cylinder rotation speed and $i = 1, 2, 3$ represents the front, top, and bottom cylinders. In addition, the lift and drag coefficients of both the system and single cylinder are recorded as follows, 

\begin{subequations}
\label{eq:whole}
\begin{equation}
C_{Di/Li}=\frac{F_{Di/Li}}{0.5\rho U^{2}DL},\label{subeq:1}
\end{equation}
\begin{equation}
    \bar{C}_{Di/Li} = \frac{1}{T}\int C_{Di/Li}(t)dt, \label{subeq:2}
\end{equation}
\begin{equation}
    \tilde{C}_{Di/Li} = \sqrt{\frac{1}{T} \int (C_{Di/Li}(t)-\bar{C}_{Di/Li})^{2}dt} ,\label{subeq:3}
\end{equation}
\end{subequations}
where with the subtext $i$, $C_{Di/Li}$ represents the drag or lift coefficient of cylinder. Specifically, $C_{D0/L0}$ denotes the system's average drag or lift coefficient, and $i = 1, 2, 3$ represents the front, top, and bottom cylinders. $\tilde{C}_{Di/Li}$ is the standard deviation of $C_{Di}$ or $C_{Li}$. $\bar{C}_{Di/Li}$ is the average value of whole $C_{Di}$ or $C_{Li}$. $F_{Di/Li}$ is the drag or lift force of corresponding cylinder sharing the same indices with relevant $C_{Di/Li}$. $\rho$ denotes the fluid density and $L$ is the unit length of cylinders. 

Before performing the DRL-based flow control, we conducted a large number of simulations for three cylinders rotating at different constant speeds to both serve as the benchmark and have a deep understanding of the problem. The rotation speed of each cylinder is selected from -5 to 5 with an incremental of 1. Therefore, in total, 1,331 combinations of different cylinder rotation speeds are simulated. 

\subsection{\label{sec:level2}Numerical Method}

The numerical solver applied in this work is based on the Boundary Data Immersion Method (BDIM) \cite{weymouth2011boundary}. It solves the viscous time-dependent Navier-Stokes equations and simulates the entire domain by combining the moving body and the ambient fluid through a kernel function. The method has quadratic convergence and has been verified for many numerical simulations for a wide range of fluid problems\cite{schlanderer2017boundary,li2022fluid}.  

The mesh configuration for this work is a rectangular Cartesian grid with a dense uniform grid near the body and in the near wake, and exponential grid stretching used in the far-field and the numerical domain of $16D \times 8D$ uses a uniform inflow, zero-gradient outflow and free-slip boundary conditions on the up and low boundaries. Furthermore, no-slip boundary conditions are employed on the circular cylinder. Mesh density is expressed in terms of grid points per chord. A uniform grid of $\delta x = \delta y = D/16$ is used for the results in this work. A detailed description of the method validation is provided in Appendix A. 

\subsection{\label{sec:level2}DRL Algorithm for Real-time Feedback Control}

We employed the twin delayed deep deterministic policy gradient (TD3) algorithm \cite{fujimoto2018addressing} to perform the DRL-based flow control. The closed-loop interaction between the \textit{fluidic pinball}, i.e., the fluid environment and the DRL agent, is shown in FIG. \ref{fig1}. The fluid environment outputs the $C_{Di}$ and $C_{Li}$, i.e., the state, from the numerical solver. The \textit{fluidic pinball} uses the rotation actuation, i.e., the action from the DRL, to alter the fluid environment. The performance of the DRL-based flow control is then evaluated using the reward $r$ illustrated as equations (2) and (3) in section IV. 

In this work, the TD3 algorithm is implemented using Python code. The Extensible Markup Language-Remote Procedure Call (XML-RPC) protocol is applied for data communication between the cross-language platforms.

\textbf{Problem formulation.} We consider infinite-horizon MDP characterized by a tuple $\left(\mathcal{S}, \mathcal{A}, \mathcal{T}, \mathcal{R}, \gamma, p_0\right)$, where $\mathcal{S} \in \mathbb{R}^n$ and $\mathcal{A} \in \mathbb{R}^m$ are continuous state and action spaces, $\mathcal{T}: \mathcal{S} \times \mathcal{A} \times \mathcal{S} \mapsto \mathbb{R}_{+}$ is the transition (dynamics) distribution, $\mathcal{R}: \mathcal{S} \times \mathcal{A} \mapsto \mathbb{R}$ denotes the reward function, $\gamma \in[0,1)$ is a discount factor, and $p_0$ is the initial state distribution. We aim to learn a mapping $\Pi_\theta: \mathcal{S} \mapsto \mathcal{A}$ with parameters $\theta$ such that discounted return $\mathbb{E}_{\Gamma \sim \Pi_\theta}\left[\sum_{t=1}^{\infty} \gamma^t r_t\right], r_t = \mathcal{R}\left(\mathbf{s}_t, \mathbf{a}_t\right)$ is maximized along a trajectory $\Gamma=\left(\mathbf{s}_0, \mathbf{a}_0, \mathbf{s}_1, \mathbf{a}_1, \ldots\right)$ following $\Pi_\theta$ by sampling an action $\mathrm{a}_t \sim \Pi_\theta\left(\cdot \mid \mathbf{s}_t\right)$ and reaching state $\mathbf{s}_{t+1} \sim \mathcal{T}\left(\cdot \mid \mathbf{s}_t, \mathbf{a}_t\right)$ at each decision step $t$.

\begin{algorithm}[H]
\caption{DRL-based flow control with TD3}
\textbf{Initialize} critic networks $Q_{\omega_1}$, $Q_{\omega_2}$\ and actor network $\pi_{\theta}$ with random parameters $\omega_1$, $\omega_2$, $\theta$ \;
    
\textbf{Initialize} target networks $\omega^{\prime} \leftarrow \omega_1, \omega_2^{\prime} \leftarrow \omega_2, \theta^{\prime} \leftarrow \theta$ \;
    
\textbf{Initialize} $done=false$, $k \in \mathbb{N}_{+}$ \;
    
\textbf{Initialize} replay buffer $\mathcal{D}$, observation buffer $\mathcal{O} \in \mathbb{R}^{k}$ \;
\begin{algorithmic}[1]
\For{\ $e=1$ \ \textbf{in} \ $N_{e}$ \  } 
  \State reset CFD environment;
  \State $i \leftarrow 0$;
  \While{$done=false$}
  \State collect the observation $o_i$; 

  \State store $o_i$ into $\mathcal{O}$, state $s_i=\mathcal{O}[i:i+k]$; 

  \State randomly generate exploration noise $\epsilon_i \sim \mathcal{N}(0,\sigma^2) $; 

  \State call an action with $\epsilon_i$: 

  \State $a_i \leftarrow \operatorname{clip}\left(\pi_\theta\left(s_i\right)+\epsilon_i,-1,1\right)$; 

  \State implement $a_i$ in the CFD environment; 
  
  \State get next state $s_{i+1}$, calculate reward $r_i$; 

  \State $i = i+1$, $s_{i} = s_{i+1}$; 

  \State store $\left(s_i, a_i, r_i, s_{i+1}\right)$ into $\mathcal{D}$; 

  \If{\textit{CFD stop}}
    \State $ done = true $;
    \EndIf
  \EndWhile
  \For{\ $j=1$ \ \textbf{in} \ $N_{j}$ \  }
  \State sample $N$ $\left(s, a, r, s^{\prime}\right)$ from $\mathcal{D}$;

  \State $\epsilon^{\prime} \sim \operatorname{clip}\left(\mathcal{N}\left(0, \tilde{\sigma}^2\right),-c, c\right)$;

  \State $\tilde{a} \leftarrow \operatorname{clip}\left(\pi_{\theta^{\prime}}\left(s^{\prime}\right)+\epsilon^{\prime},-1,1\right)$;

  \State $y \leftarrow r+\gamma \min _{i=1,2} Q_{\omega_i^{\prime}}\left(s^{\prime}, \tilde{a}\right)$;

  \State $L_{\omega_i} \leftarrow N^{-1} \sum\left(y-Q_{\omega_i}(s, a)\right)^2, i=1,2$;

  \State update $\omega_i$ with the loss function $L_{\omega_i}, i=1, 2$;

  \If{$e \bmod t$}
  
  \State $L_\theta \leftarrow-N^{-1} \sum Q_{\omega_1}\left(s, \pi_\theta(s)\right)$;

  \State update $\theta$ with the loss function $L_\theta$;

  \State update the target networks: 

  \State $\omega_i^{\prime} \leftarrow \tau \omega_i+(1-\tau) \omega_i^{\prime}, i=1,2$;

  \State $\theta^{\prime} \leftarrow \tau \theta+(1-\tau) \theta^{\prime}$.
  \EndIf
\EndFor
\EndFor

\end{algorithmic}
\end{algorithm}

TD3 algorithm aims to estimate an optimal state-action value function $Q^*: \mathcal{S} \times \mathcal{A} \mapsto \mathbb{R}$ using a value function $Q_\omega(\mathbf{s}, \mathbf{a}) \approx Q^*(\mathbf{s}, \mathbf{a})=\max _{\mathbf{a}^{\prime}} \mathbb{E}[\mathcal{R}(\mathbf{s}, \mathbf{a})+$ $\left.\gamma Q^*\left(\mathbf{s}^{\prime}, \mathbf{a}^{\prime}\right)\right] \forall \mathbf{s} \in \mathcal{S}$ where $\mathbf{s}^{\prime}, \mathbf{a}^{\prime}$ is the state and action at the following step, and $\omega$ parameterizes the function. While $Q^*$ is generally unknown, it can be approximated by repeatedly fitting $Q_\omega$ using the update rule:
$$
\omega^{k+1} \leftarrow \arg \min _\omega \mathbb{E}_{\left(\mathbf{s}, \mathbf{a}, \mathbf{s}^{\prime}\right) \sim \mathcal{B}}\left\|Q_\omega(\mathbf{s}, \mathbf{a})-y\right\|_2^2
$$
where the $Q$-target $y=\mathcal{R}(\mathbf{s}, \mathbf{a})+\gamma \max _{\mathbf{a}^{\prime}} Q_{\omega^{-}}\left(\mathbf{s}^{\prime}, \mathbf{a}^{\prime}\right), \mathcal{B}$ is a replay buffer that is iteratively grown as new data is collected, and $\omega^{-}$is a slow-moving average of the online parameters $\omega$ updated with the rule $\omega_{k+1}^{-} \longleftarrow(1-\zeta) \omega_k^{-}+$ $\zeta \omega_k$ using a constant coefficient $\zeta \in[0,1)$. In DRL, generally, $\Pi$ is typically a policy parameterized by a neural network that learns to approximate $\Pi_\theta(\cdot \mid \mathrm{s}) \approx$ $\arg \max _{\mathbf{a}} \mathbb{E}\left[Q_\omega(\mathbf{s}, \mathbf{a})\right] \forall \mathbf{s} \in \mathcal{S}$, i.e., the globally optimal policy.


\textbf{DRL-based flow control with TD3} (Algorithm 1) is a deterministic strategy reinforcement learning algorithm suitable for continuous action space. It greatly improves the learning speed and performance of previous algorithms in numerous challenging tasks in the continuous control setting, which exceeds the performance of numerous state-of-the-art DRL algorithms \cite{fujimoto2018addressing}. Such merit of TD3 determines its vital role in this work. All the neural networks are feedforward neural networks with two hidden layers, each of width 256 in this work. The discount factor $\gamma$ is set as 0.99, and the policy exploration Gaussian noise $\sigma$ is set as 0.1 in the simulation. We use the Adam optimizer with learning rate $10^{-4}$ to update $\omega_{i}$ and $\theta$. $N_{j}$ is set to 1000, and the batch size is set to 512. The regularization noise $\tilde{\sigma }$ is set to 0.2, and $c$ is set to 0.5. The target networks are updated every $d =2$ iterations, and the soft updating rate $\tau$ is set to 0.005. 

Unlike traditional tasks of DRL, flow control is not a standard MDP, as a long-time flow field lasting many steps can impact the next state. Two techniques are presented to solve the problem of employing the DRL on such flow control tasks. Firstly, unlike the general TD3 algorithm, which learns the policy online, the tuned TD3 algorithm in this work transfers the online training into the offline training mode. In each training, the agent can randomly sample data from the buffer filled with the trajectory of the whole flow control process, which means that the agent can access the relative overall flow field feature from start to end. Secondly, states are encoded by current $C_{Di}^{t}$ and $C_{Li}^{t}$ combining with them in many previous steps as $s_{t}=(C_{Li}^{t-n}, C_{Di}^{t-n}, C_{Li}^{t-n+1}, C_{Di}^{t-n+1},\ldots,  C_{Li}^{t}, C_{Di}^{t})$, where $t$ is the current step and $n$ is the memory steps defined according to the tasks. $n$ is set as 32 for this work. 

\begin{figure*}
    \centering
        \includegraphics[width=1\textwidth]{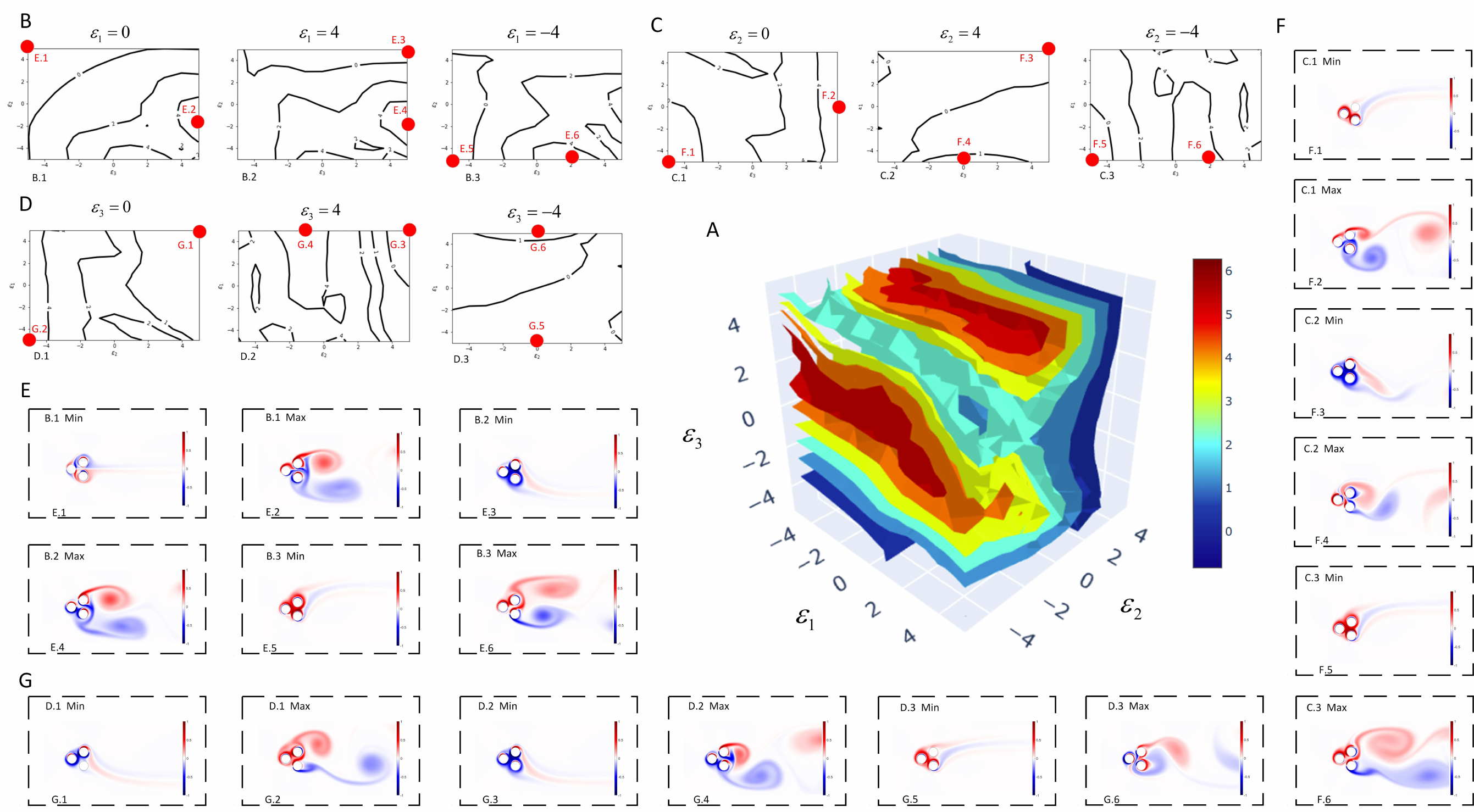}
        \caption{\textbf{Average \(C_{D0}\) of \textit{fluidic pinball} system in the three-dimensional parametric space using open-loop control with constant rotation speeds.} Isosurface, contours, and vorticities of the $C_{D0}$: (A) Point at each position is the \(C_{D0}\) corresponding to the rotations of cylinders of 1,331 simulations $(C_{D0}\sim f(\varepsilon_{1},\varepsilon_{2},\varepsilon_{3}))$. Red and light colors mean \(C_{D0}\) is higher, and blue and dark colors mean \(C_{D0}\) is smaller and negative. (B),(C),(D) Contours of sections on three coordinate axes of subfigure A at the value equal to 0, 4, -4, where $C_{D0}\sim f(\varepsilon_{i},\varepsilon_{j}), i\in[1,2,3]\neq j\in[1,2,3]$. B.1, B.2, B.3 are \(C_{D0}\) of $\varepsilon_{1}=0,4,-4$; C.1, C.2, C.3 are \(C_{D0}\) of $\varepsilon_{2}=0,4,-4$; D.1, D.2, D.3 are $\varepsilon_{3}=0,4,-4$ respectively. One of the fixed rotations are marked on top of contours. (E),(F),(G) Snapshot vorticities extracted of red spots from subfigures B, C and D, where rotation speeds of three cylinders are marked for each case. Their corresponding labels can be found near red spots on contours. There are also the simplified tabs on vorticities, with max and min which means its extremum in the contour. Wake patterns of \textit{fluidic pinball} depend on the patterns of all cylinders' rotation leading to related drag variation, and \(C_{D0}\) of constant rotations on rear cylinders show the symmetrical relationship as their symmetric positions along the center-line of the incoming flow.}
        \label{fig:cd_all}
\end{figure*}

\section{\label{sec:ccrs}Results of Cylinders with Constant Rotation Speed}

In order to understand better and explore the trend of forces on the \textit{fluidic pinball} with the adjustment of $\varepsilon_{i}$, we analyzed the 1,331 simulations' results of open-loop control with constant rotation speed based on their forces and vortical wakes in detail. Meanwhile, these results from brute-force searching are valuable references for the DRL-based feedback control to verify the performance of the DRL algorithm. 

\begin{figure*}
    \centering
        \includegraphics[width=1\textwidth]{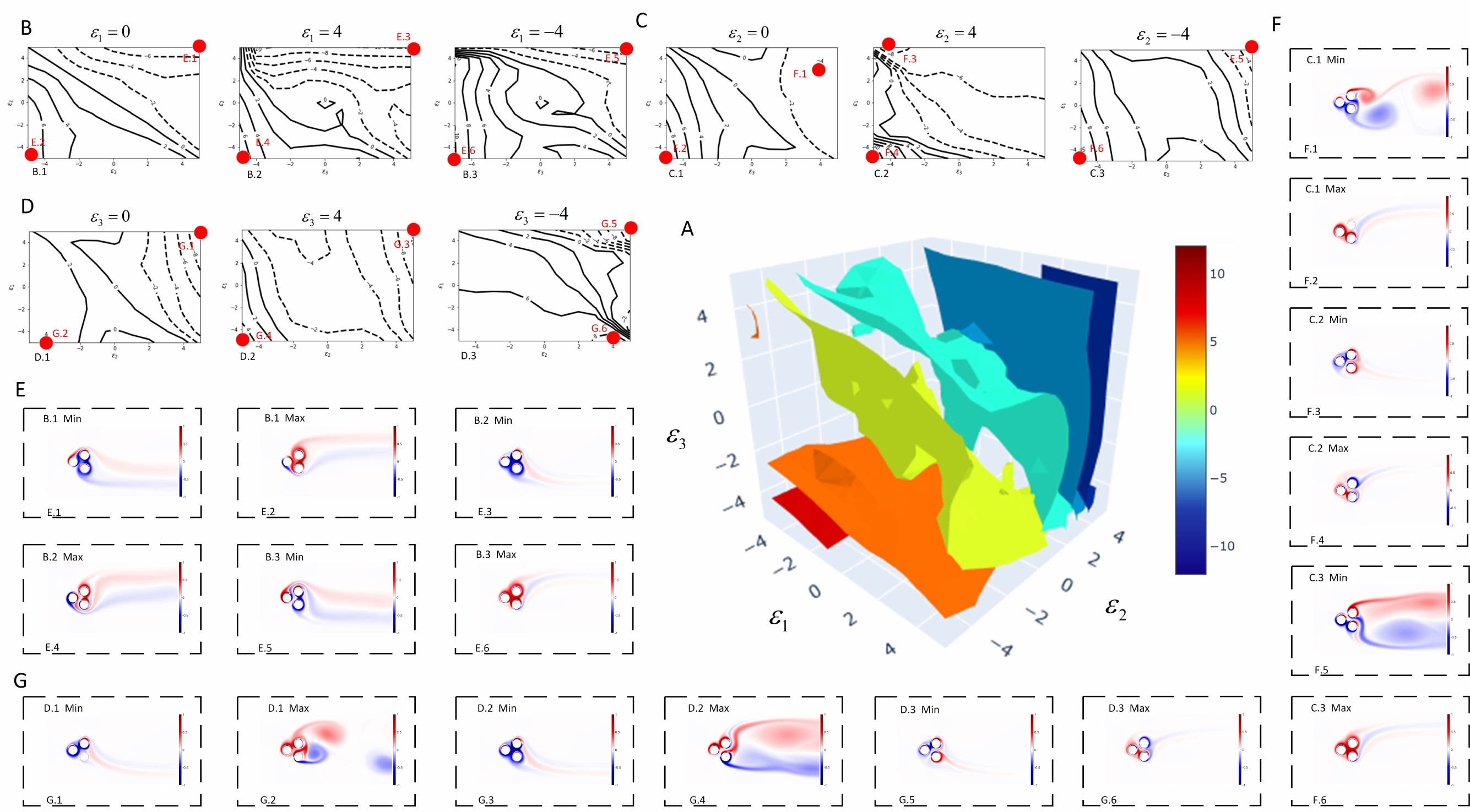}
        \caption{\textbf{Average \(C_{L0}\) of \textit{fluidic pinball} system in the three-dimensional parametric space using open-loop control with constant rotation speeds.} Isosurface, contours, and vorticities of the \(C_{L0}\): (A) Point at each position is the \(C_{L0}\) corresponding to the rotations of cylinders of 1,331 simulations $(C_{L0}\sim f(\varepsilon_{1},\varepsilon_{2},\varepsilon_{3}))$. Red and light colors mean \(C_{L0}\) is higher, and blue and dark colors mean \(C_{L0}\) is smaller and negative. (B),(C),(D) Contours of sections on three coordinate axes of subfigure A at the value equal to 0, 4, -4, where $C_{L0}\sim f(\varepsilon_{i},\varepsilon_{j}), i\in[1,2,3]\neq j\in[1,2,3]$. B.1, B.2, B.3 are \(C_{L0}\) of $\varepsilon_{1}=0,4,-4$; C.1, C.2, C.3 are \(C_{L0}\) of $\varepsilon_{2}=0,4,-4$; D.1, D.2, D.3 are $\varepsilon_{3}=0,4,-4$ respectively. One of the fixed rotations are marked on top of contours. (E),(F),(G) Snapshot vorticities extracted of red spots from subfigures B, C, and D, where rotation speeds of three cylinders are marked for each case. Their corresponding labels can be found near red spots on contours. There are also the simplified tabs on vorticities, with max and min which means its extremum in the contour. \(C_{L0}\) show a relatively simple and monotonic variation pattern, which has a type of symmetrical and reverse relationship in cases of constant rotations on rear cylinders.}
        \label{fig:clmean_all}
\end{figure*}

\begin{figure*}
    \centering
        \includegraphics[width=1\textwidth]{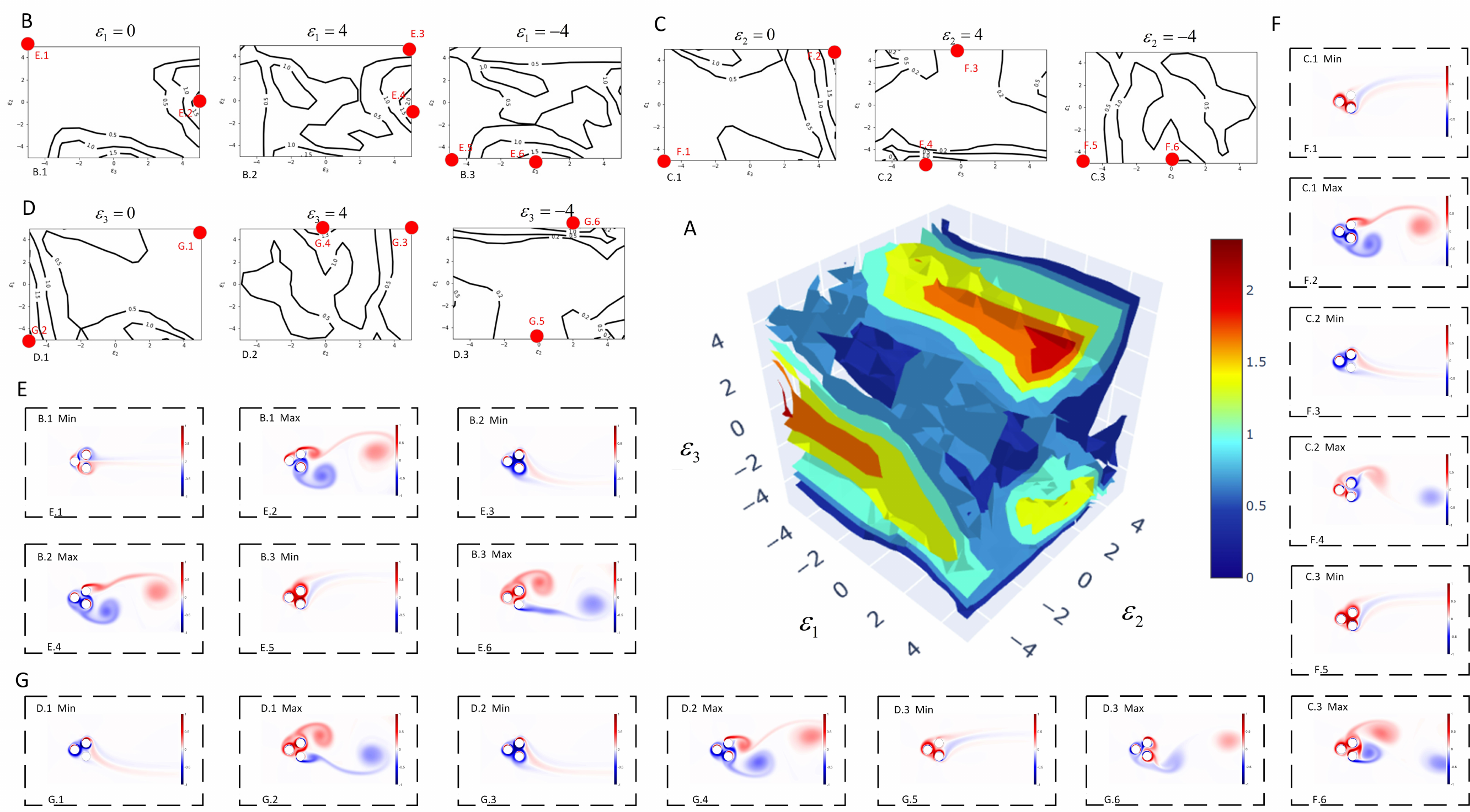}
        \caption{\textbf{Average \(\tilde{C}_{L0}\) of \textit{fluidic pinball} system in the three-dimensional parametric space using open-loop control with constant rotation speeds.} Isosurface, contours, and vorticities of the \(\tilde{C}_{L0}\): (A) Point at each position is the \(\tilde{C}_{L0}\) corresponding to the rotations of cylinders of ,1331 simulations $(\tilde{C}_{L0}\sim f(\varepsilon_{1},\varepsilon_{2},\varepsilon_{3}))$. Red and light colors mean \(\tilde{C}_{L0}\) is higher, and blue and dark colors mean \(\tilde{C}_{L0}\) is smaller and negative. (B),(C),(D) Contours of sections on three coordinate axes of subfigure A at the value equal to 0, 4, -4, where $\tilde{C}_{L0}\sim f(\varepsilon_{i},var\epsilon_{j}), i\in[1,2,3]\neq j\in[1,2,3]$. B.1, B.2, B.3 are \(\tilde{C}_{L0}\) of $\varepsilon_{1}=0,4,-4$; C.1, C.2, C.3 are \(\tilde{C}_{L0}\) of $\varepsilon_{2}=0,4,-4$; D.1, D.2, D.3 are $\varepsilon_{3}=0,4,-4$ respectively. One of the fixed rotations are marked on top of contours. (E),(F),(G) Snapshot vorticities extracted of red spots from subfigures B, C, and D, where rotation speeds of three cylinders are marked for each case. Their corresponding labels can be found near red spots on contours. There are also the simplified tabs on vorticities, with max and min which means its extremum in the contour. Narrow wakes reduce \(\tilde{C}_{L0}\) and the cases with minimum \(\tilde{C}_{L0}\) has narrow wake while cases with maximum \(\tilde{C}_{L0}\) has wide wake.}
        \label{fig:clstd_all}
\end{figure*}

FIG. \ref{fig:cd_all}, FIG. \ref{fig:clmean_all} and FIG. \ref{fig:clstd_all} are the average of the drag coefficient \(C_{D0}\), the average of the lift coefficient \(C_{L0}\) and the standard deviation of the lift coefficient \(\tilde{C}_{L0}\) of \textit{fluidic pinball} system with respect to $(\varepsilon_1, \varepsilon_2, \varepsilon_3)$ separately. Subfigure A is the isosurface of the coefficients in the three-dimensional parametric space corresponding to the rotations of cylinders of 1,331 simulations; Subfigures B, C, and D are the contours of sections on three coordinate axes of subfigure A at the value equal to 0, 4, -4, where the coefficients $C_{Di/Li}\sim f(\varepsilon_{i}, \varepsilon_{j}), i\in[1,2,3]\neq j\in[1,2,3]$; Subfigures E, F, and G are snapshot vorticities extracted of red spots from subfigures B, C, D respectively.

\subsection{\label{sec:force}Force distribution}

Our main observations can be summarized as follows:

\textbf{For average \(C_{D0}\)}:
\begin{enumerate}
    \item Along with the increasing of \(\varepsilon_{3}\), \(C_{D0}\) is increasing firstly and then decreasing when \(\varepsilon_{2}<-2\). When \(\varepsilon_{2}>-2\), \(C_{D0}\) is increasing along with the increasing of \(\varepsilon_{3}\).
    \item Along with the increasing of \(\varepsilon_{2}\), \(C_{D0}\) is increasing firstly and then decreasing when \(\varepsilon_{3}>2\). When \(\varepsilon_{3}<2\), \(C_{D0}\) is decreasing along with the increasing of \(\varepsilon_{3}\).
    \item The changing of \(\varepsilon_{1}\) does not have much impact on \(C_{D0}\).
    \item There are two maximum regions. When \(\varepsilon_{2}\in(-3,-1)\) and \(\varepsilon_{3}\) is near 5, \(C_{D0}\) has the maximum region large more than 6. When \(\varepsilon_{3}\in(1,3)\) and \(\varepsilon_{2}\) is near -5,  \(C_{D0}\) is large more than 6 as well.
    \item From above maximum regions to the periphery, \(C_{D0}\) decreases gradually, achieving the broad minimum regions less than 0.
\end{enumerate}

\textbf{For average \(C_{L0}\)}:
\begin{enumerate}
    \item \(C_{L0}\) is decreasing along with the increasing of \(\varepsilon_{2}\)
    \item When \(\varepsilon_{2}<1\), \(C_{L0}\) is decreasing along with the increasing of \(\varepsilon_{3}\). When \(\varepsilon_{2}>2\), \(C_{L0}\) is always less than 0, shown in FIG. 3A.
    \item The maximum region about 10 is located at \(\varepsilon_{1}<-2\) and \(\varepsilon_{3}\) near -5.
    \item The minimum region about -10 is at \(\varepsilon_{1}>2\) and \(\varepsilon_{2}\) near 5.
\end{enumerate}

\textbf{For average \(\tilde{C}_{L0}\)}:
\begin{enumerate}
    \item Most of the region is less than 1 shown in FIG. 4A.
    \item When \(\varepsilon_{3}>3\), \(\tilde{C}_{L0}\) is increasing along with the increasing of \(\varepsilon_{3}\), and increasing firstly and then decreasing along with the increasing of \(\varepsilon_{2}\).
    \item When \(\varepsilon_{2}<-3\), \(\tilde{C}_{L0}\) is increasing along with the decreasing of \(\varepsilon_{2}\), and increasing firstly and then decreasing along with the decreasing of \(\varepsilon_{3}\).
    \item There are two maximum regions. \(\varepsilon_{1}>4\), \(\varepsilon_{2}\in(-2,0)\), \(varepsilon_{3}\) near 5, and \(\varepsilon_{3}\in(0,2)\) \(\varepsilon_{2}\) and \(\varepsilon_{1}\) near -5 have the maximum \(\tilde{C}_{L0}\) more than 2.
\end{enumerate}

\begin{figure}
    \centering
    \includegraphics[width=0.5\textwidth]{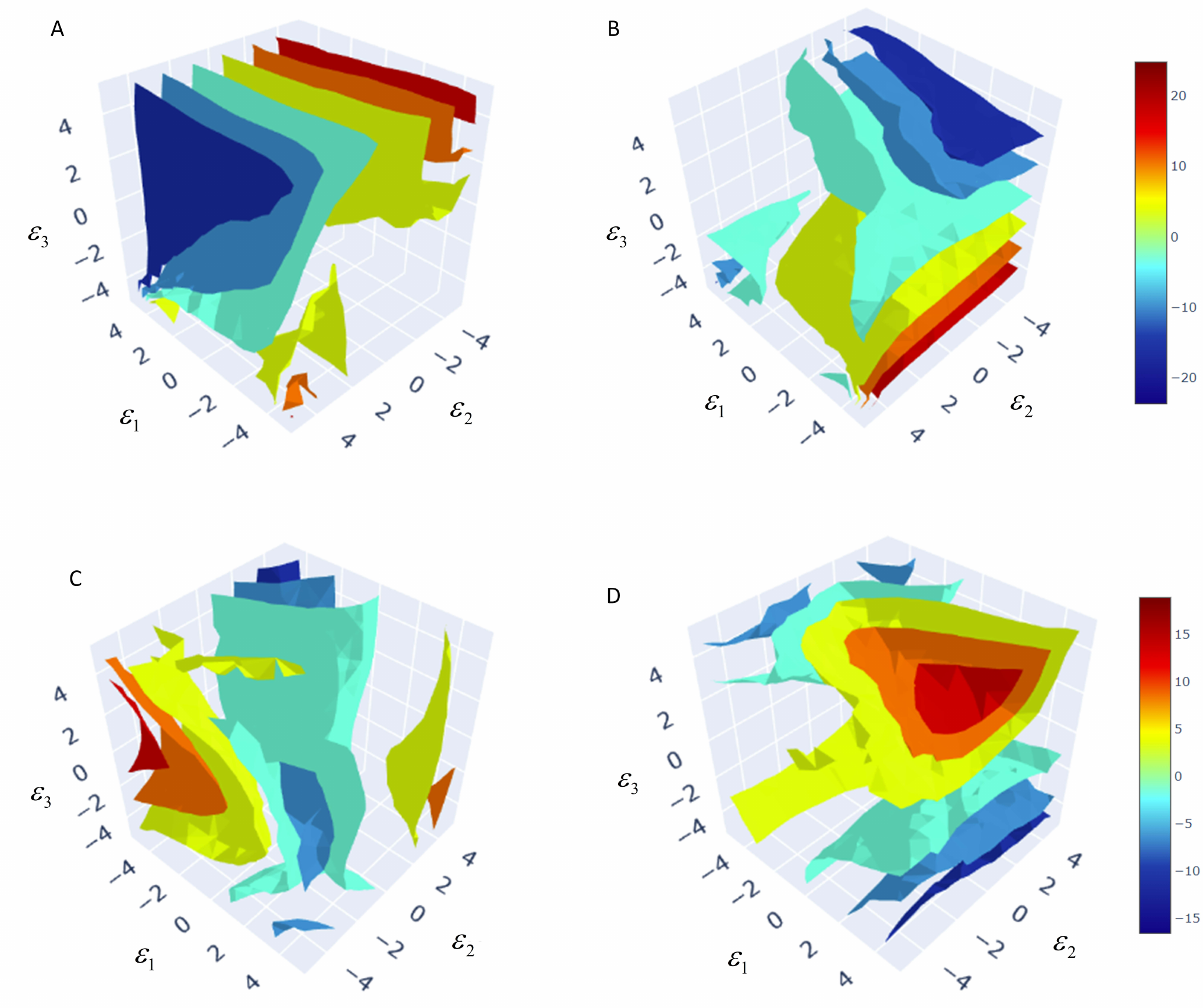}
    \caption{\textbf{\(C_{Li}\) and \(C_{Di}\) of the top and bottom cylinders.} \(C_{Li}\) and \(C_{Di}\) on two rear cylinders in the three-dimensional parametric space. Values in isosurfaces are the coefficients on a single cylinder corresponding to the rotations of three cylinders, which are functions of \(\varepsilon_{1}\),\(\varepsilon_{2}\) and \(\varepsilon_{3}\): (A) Points in isosurface are \(C_{L2}\) on the top cylinder; where red and light colors mean values are higher, and blue and dark colors mean values are smaller and negative. (B) Points in the isosurface are \(C_{L3}\) on the bottom cylinder corresponding to the rotations of cylinders. (C) Points in isosurface are \(C_{D2}\) on the top cylinder, where red and light colors mean values are higher, and blue and dark colors mean values are smaller and negative. (D) Points in the isosurface are \(C_{D3}\) on the bottom cylinder corresponding to the rotations of cylinders. The \(C_{Li}\) on the bottom and top cylinders have strong monotonicity along with the speeds' variations of their own cylinders, while the variation patterns of \(C_{Di}\) on them are more complex. 
}
    \label{fig:rearCyl}
\end{figure}

Moreover, \(C_{D2}\) and \(C_{D3}\) of the rear cylinders show a symmetrical relationship with each other. In addition, \(C_{L2}\) and \(C_{L3}\) show a symmetrical and reverse relationship from each other. Moreover, the \(C_{L2}\) of the top cylinder and \(C_{L3}\) of the bottom cylinder have strong monotonicity along with the speeds' variations of their own cylinders, while the variation patterns of \(C_{Di}\) on them are more complex shown in FIG. \ref{fig:rearCyl}.

\subsection{\label{sec:level2}Vortical wake analysis}

From the force distribution, we observe that different rotation combinations lead to various patterns of vortical wakes and will cause different forces on \textit{fluidic pinball}. To get an insight into the patterns of  wakes impacts the force distribution, we selected some key vorticities in FIG. \ref{fig:cd_all}, FIG. \ref{fig:clmean_all} and FIG. \ref{fig:clstd_all} as representative to demonstrate the complex vortical wakes behind the \textit{fludic pinball} as the function of $\varepsilon_{1},\varepsilon_{2},\varepsilon_{3}$.

\textbf{For average \(C_{D0}\)}

In FIG. \ref{fig:cd_all}, E.2, E.4, E.6, F.2, F.4, F.6, G.2, G.4, and G.6 show the complex vortex pattern in the near wake and Karman vortex street in far wake. The rotation speeds of three cylinders in vector form are (0,-2,5), (4,-2,5), (-4,-5,2), (0,0,5), (-5,4,0), (-5,-4,2), (-5,-5,0), (5,-1,4), (5,0,-4) respectively. If the rotation of cylinders is divided into the fix, clockwise, and counterclockwise rotations, two rear cylinders in these 9 cases do not have the same performance leading to this kind of complex vortex pattern. On the other hand, the front cylinders perform all rotation patterns, which means it does not have a decisive influence on this wake. They have the maximum $C_{D0}$ in their contours as the wakes lead to lower pressure behind the \textit{fluidic pinball}. There are complex and chaotic vortices in near wakes improving the velocity after \textit{fluidic pinball} and reducing the pressure as the Bernoulli effect. Then $C_{D0}$ becomes positive and similar as the difference of pressure between upstream and downstream of the \textit{fluidic pinball} system. 

On the other hand, E.1, E.3, E.5, F.1, F.3, F.5, G.1, G.3, and G.5 show a smooth wake after the \textit{fluidic pinball}. The rotation speeds are (0,5,-5), (4,5,5), (-4,-5,-5), (-5,0,-5), (5,4,5), (-5,-4,-5), (5,5,0), (5,5,4), (-5,0,-4) respectively. In these cases, two rear cylinders do not have the reverse rotations, except for E.1, where rear cylinders rotate in equal and reverse high velocities (a boat-tailing mechanism), leading to a relatively small $C_{D0}$. The front cylinder, in most cases, rotates in the same direction as the rear except for the fixed front one in E.1, which leads to a slight $C_{D0}$ on the \textit{fluidic pinball}. The smooth and narrow wakes reduce the flow velocities after the \textit{fluidic pinball}. Thus, higher pressure after the system leads to a negative $C_{D0}$.

With the comparison between E.3 (-4,5,5) and E.4 (-4,-2,5) in FIG. \ref{fig:cd_all}, only the rotation of the top cylinder has the change from clockwise to counterclockwise, which transfers the wake from a pattern of wake deflecting to a downward direction to alternating shedding. As the rear cylinders in E.4 have reverse rotations to the inner, their shears bleed from the gap leading to the complex vortex and high $C_{D0}$. After that, between G.5 (-5,0,-4) and G.6 (5,0,-4) in FIG. \ref{fig:cd_all}, only the front cylinder's rotation has changed from counterclockwise to clockwise. As the front cylinder in G.5 has a counterclockwise rotation, its shears deflect up and then extend to wake. Some smooth and independent shears in the wake decrease the flow velocity and improve the pressure leading to the negative $C_{D0}$. As the front and bottom cylinders in G.6 have reverse clockwise rotations, the pressure after the \textit{fluidic pinball} is reduced, leading to the positive $C_{D0}$. Similarly, between F.5 (-5,-4,-5) and F.6 (-5,-4,2), only the rotation of the bottom cylinder has changed from counterclockwise to clockwise, which transfers the wake from the pattern of wake deflecting to up to alternating shedding. $C_{D0}$ is more significant than 4 in F.6, while $C_{D0}$ is smaller than 0 in F.5 as the change of rotation of the top cylinder.

We can also discover that F.4 (-5,4,0) and G.6 (5,0,-4) have the same $C_{D0}$ near 1. Meanwhile, F.5 (-5,-4,-5) and G.3 (5,5,4) have the same $C_{D0}$ of less than 0. However, the patterns of their wakes are different. The wake of F.5 deflects to the upside while that of G.3 deflects to the down. The wakes of F.4 and G.6 have a symmetric relationship. Except for the vorticities having a symmetric relationship, the contours like C.2 and D.3 are also symmetric. C.1 and D.1, C.2 and D.3, C.3 and D.2 in FIG. \ref{fig:cd_all} have 180-degree rotation symmetry. The drags of inverse rotation speeds on three cylinders with the swap of two rear cylinders are equal.

\textbf{For average \(C_{L0}\)}

In FIG. \ref{fig:clmean_all}, E.2, E.4, E.6, F.2, F.4, F.6, G.6 show the wakes deflecting to the up position. The rotation speeds of three cylinders in vector form are (0,-5,-5), (4,-5,-5), (-4,-5,-5), (-5,0,-5), (-5,4,-5), (-5,-4,-5), (-5,4,-4) respectively. There is always a rear cylinder (bottom one or top one) or both of them rotating in the counterclockwise direction so that the shears can be controlled to deflect up. They have the maximum $C_{L0}$ in their contours as the wakes lead to a difference in pressure. As the wake accelerates the flow velocity, the pressure on the upside is lower, caused by higher flow velocity as the Bernoulli effect if the wake deflects into the up direction. Thus, the lift is positive and from down to up in a vertical direction. There are two cases G.2 and G.4 with rotation (-5,-4,0) and (-5,-5,4) also have the maximum $C_{L0}$ in their group, which do not have the up deflecting wake. Drag in G.2 and G.4 is near 4 with the relative complex wake, but the wake also has a deflecting trend.

On the other hand, E.1, E.3, E.5, F.3, G.1, G.3, and G.5 show the wake after the \textit{fluidic pinball} deflecting to the down direction. The rotation speeds are (0,5,5), (4,5,5), (-4,5,5), (5,4,-4), (5,5,0), (5,5,4), and (5,5,-4), respectively. There is always a rear cylinder (bottom one or top one) or both of them rotating in the clockwise direction so that the shears can be controlled to deflect down. Thus, the pressure on the downside is lower, caused by higher flow velocity leading to the negative lifts of \textit{fluidic pinball}. They have the minimum and negative $C_{L0}$ in their contours. There are two cases, F.1 and F.5, with rotation (3,0,4) and (5,-4,5) also having the negative $C_{L0}$ in their groups, which has a relatively complex pattern in near wake after cylinders where the shear of the top cylinder deflects to down and merges with shears of front and bottom cylinders leading the wake to have a down deflecting trend as well.

With the comparison between F.3 (5,4,-4) and G.6 (-5,4,-4) in FIG. \ref{fig:clmean_all}, only the rotation of the front cylinder has the change from clockwise to counterclockwise, which transfers the wake from down deflection to up deflection. The reverse rotation of the front cylinder also leads to inverse deflection and inverse $C_{L0}$. After that, between F.4 (-5,4,-5) and G.5 (5,5,-4), except for the reverse rotation on the front cylinder, there are slight tunes for the rear cylinders. The $C_{L0}$ transfers from the positive to negative as the deflective changes of wakes. Similarly, the difference between F.5 (5,-4,5) and G.4 (-5,-5,4) also sees the same changes of wakes and $C_{L0}$ caused by the reverse rotation of the front cylinder, although its rotation is relatively small in the two cases. Then, the rotation direction of the front cylinder has a decisive influence on the value of $C_{L0}$ on \textit{fluidic pinball}. 

\begin{figure*}
  \centerline{\includegraphics[width=1\textwidth]{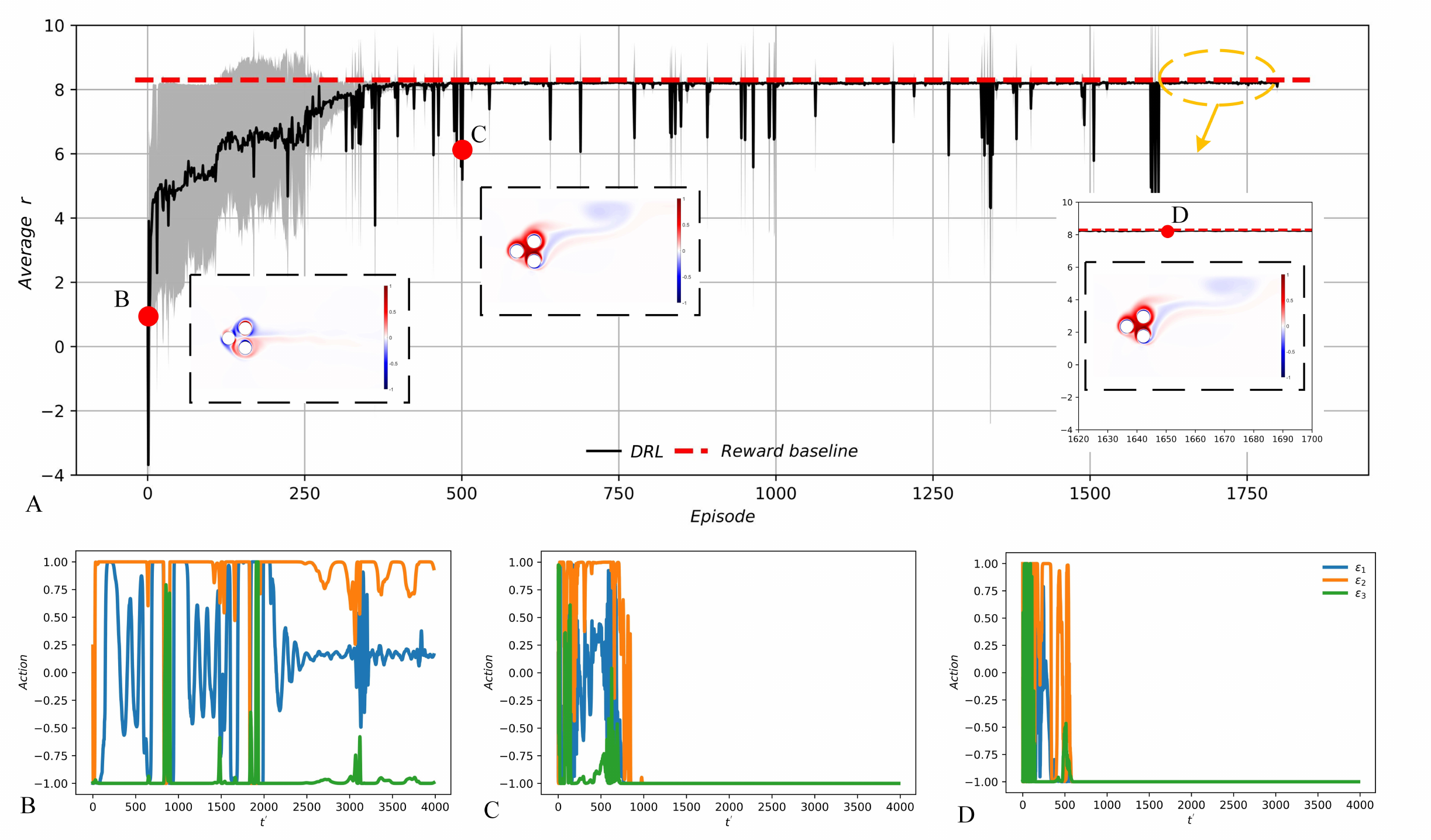}}
  \caption{\textbf{Training process of DRL-based flow control for forces extremum.} The training process represented by the average $r$ against the episode number of four different agents, the corresponding actions evaluated by agents, and vortical wake visualization of various training stages, where $t^{'}=\frac{tD}{U}$. (A) The changing of average $r$ and the vortical wakes of the last time step at three training stages. The shown line here is mean value and the shade area is corresponding variance. (B), (C) and (D) The changing of generated actions in evaluation at three training stages sharing the same legend.}
\label{extre}
\end{figure*}

From the above three comparisons, we can discover that F.4 (-5,4,-5) and G.5 (5,5,-4) have the same absolute value of $C_{L0}$ of more than 16 and reverse direction. Meanwhile, F.2 (-5,0,-5) and G.1 (5,5,0) have the same absolute value of $C_{L0}$ of more than 8 in the reverse direction as well. However, the patterns of their wakes are different. The Wake of F.4 deflects up while that of G.5 deflects down, the wakes of F.2 deflect up, and G.1 deflects to the down direction, which has a symmetric relationship about the horizontal center-line leading to the inverse directions and the same absolute value of their $C_{L0}$ respectively. Except for the vorticities having a symmetric relationship, the contours like C.2 and D.3 are also symmetric. C.1 and D.1, C.2 and D.3, C.3 and D.2 in FIG. \ref{fig:clmean_all} have 180-degree rotation and reverse symmetry different from the $C_{D0}$. The $C_{L0}$ of inverse rotation speeds on three cylinders with the swap of two rear cylinders are reversed.

\textbf{For average \(\tilde{C}_{L0}\)}

FIG. \ref{fig:clstd_all}, E.2, E.4, E.6, F.2, F.4, F.6, G.2, G.4, G.6 show the wide wakes alternating periodic shedding similar to Karman vortex street. The rotation speeds of three cylinders in vector form are (0,0,5), (4,0,5), (-4,-5,0), (5,0,5), (-5,4,-2), (-5,-4,0), (-5,-5,0), (5,0,4), (5,2,-4) respectively, which have large \(\tilde{C}_{L0}\). The coordinative rotations of cylinders lead to the vast wakes where the up vortex and down vortex alternating shed. They have the maximum \(\tilde{C}_{L0}\) in their contours as the alternating periodic vortices lead to large fluctuations in overall lifts.

On the other hand, E.1, E.3, E.5, F.1, F.3, F.5, G.1, G.3, and G.5 show the smooth and narrow wake after the \textit{fluidic pinball} deflecting to any direction. The rotation speeds are (0,5,-5), (4,5,5), (-4,-5,-5), (-5,0,-5), (5,4,0), (-5,-4,-5), (5,5,0), (5,5,4), (-5,0,-4) respectively. They have the minimum \(\tilde{C}_{L0}\) in their contours. Whether the wakes deflect in up or down directions, the smooth wakes do not cause large fluctuations of lift on \textit{fluidic pinball}, leading a very small \(\tilde{C}_{L0}\) (near 0) in general.

In FIG. \ref{fig:clstd_all}, with the comparison between F.2 (5,0,5) and G.2 (-5,-5,0), they have the same \(\tilde{C}_{L0}\) near 2 but have different wakes, which are symmetric about the horizontal center-line. \(\tilde{C}_{L0}\) between F.3 (5,4,0) and G.5 (-5,0,-4) are equally less than 0.2, where the wakes are also symmetric about the horizontal center-line deflecting to up or down directions. Similarly, for F.6 (-5,4,0) and G.4 (5,0,4), their \(\tilde{C}_{L0}\) are same near 1.5. Their wakes are symmetric about the horizontal center-line, with alternating shedding vortices. From the above comparisons, we can discover that some cases using different action pairs with the symmetric wakes have the same \(\tilde{C}_{L0}\). Except for the vorticities having a symmetric relationship, the contours like C.2 and D.3 are also symmetric. C.1 and D.1, C.2 and D.3, C.3 and D.2 in FIG. \ref{fig:clstd_all} have 180-degree rotation symmetry. \(\tilde{C}_{L0}\) of inverse rotation speeds on three cylinders with the swap of two rear cylinders are equal.

\begin{figure*}
  \centerline{\includegraphics[width=1\textwidth]{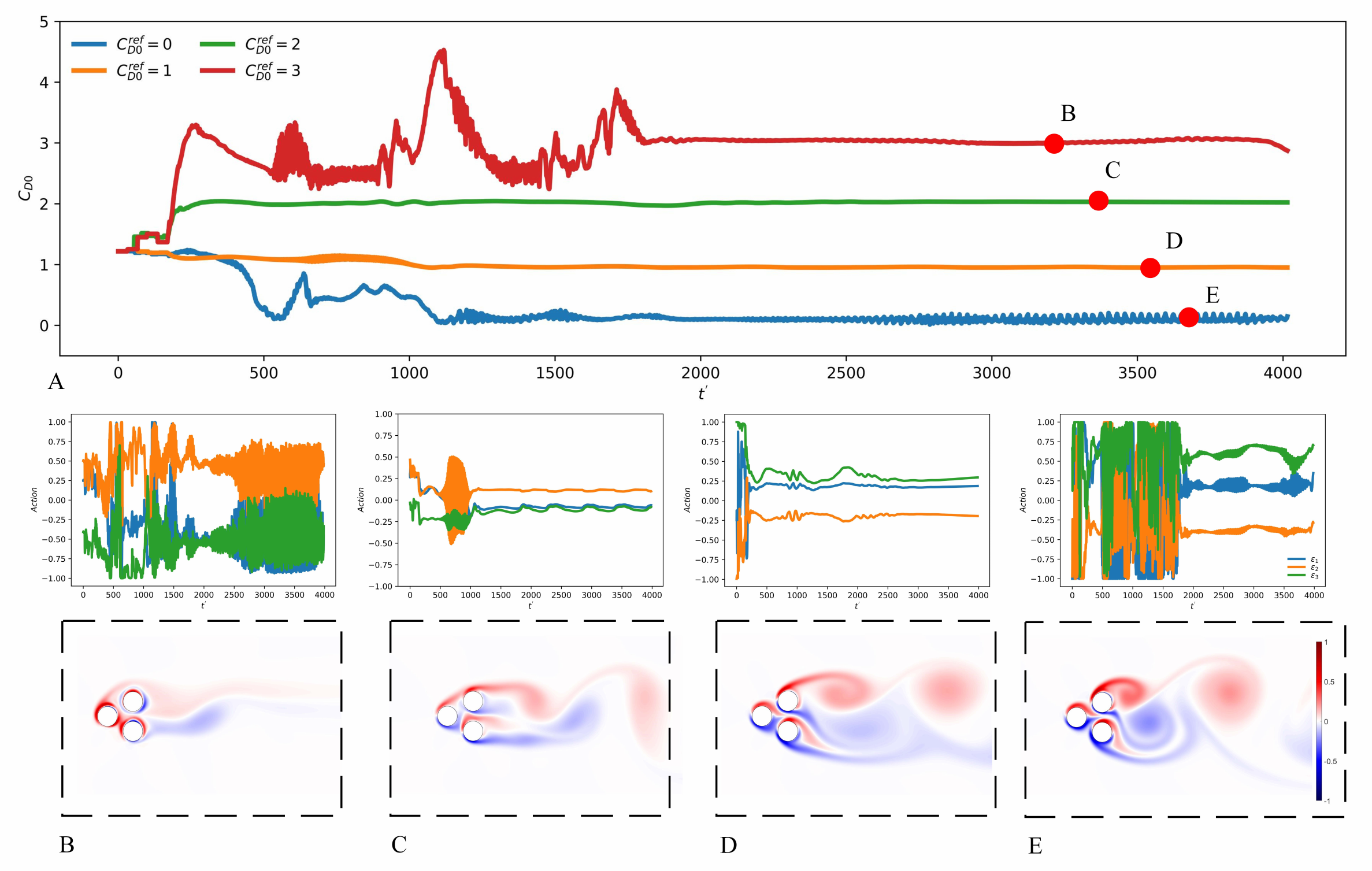}}
  \caption{\textbf{Result of DRL-based flow control for forces tracking.} The $C_{D0}$ in the time domain for four force tracking problems, vortical wake visualization of the last time step after convergence, and the corresponding actions generated by agents. (A) The changing of $C_{D0}$ that retain at 0, 1, 2, and 3 sharing the same legend. (B), (C), (D), (E) The vortical wake extracted in the last time step from four tests, respectively, sharing the same color bar in subfigure A and the corresponding actions.}
\label{track}
\end{figure*}

\section{\label{sec:level1}Results of DRL-based Feedback Control}
\subsection{\label{sec:level2}DRL-based Force Extremum and Tracking}

In the first task, to demonstrate the validity of the DRL algorithm applied in the current work, we attempt to compare the DRL optimization result with that found in brute-force searching by minimizing the drag on the cylinder while maximizing the lift. The results of the $C_{Di}$, $C_{Li}$ and actions are plotted in FIG. \ref{extre}, while the setup of the reward $r$ in the first task are as follows:

\begin{equation}
    r=-sgn(C_{Di})(\left | C_{Di} \right |)^{1/2}+sgn(C_{Li})(\left | C_{Li} \right |)^{1/2},
\end{equation}

where the first term denotes the reward for minimizing drag (maximizing thrust) on the front cylinder, and the second term is the reward for maximum and positive direction lift on the bottom cylinder. We reduce the drag of the front cylinder and enhance the lift on the bottom cylinder simultaneously, demonstrating the ability of DRL to find force extremum. The DRL-based method can control the drag and lift to the minimum and maximum values, respectively, like that searched by open-loop control with the same reward values.

DRL-based force control results show that $C_{Di}$ and $C_{Li}$ can achieve steady extremum. The training process is shown in FIG. \ref{extre} with 6A, the average reward (black line) that converges is calculated by Eq. (2), and the vortical wakes in the last time step of the three training stages are shown near the curve. 6B, 6C, and 6D are the corresponding actions for three circular cylinders normalized in the range of $(-1,1)$. FIG. 6A shows that the reward increases from the initial value to the convergence position near 8.3, comparable with the reward-maximum case calculated from the brute-force searching (red line). Such a result indicates that the DRL-based control can reach as good a result as the exhaustive brute-force searching. The actions in the initial stage in 6B have a different pattern compared with the optimal actions in 6D, where three cylinders rotate in a counterclockwise direction with the fastest normalized speed (-1.00,-1.00,-1.00), which is the case with minimum $C_{D0}$ on the front cylinder and maximum $C_{L0}$ on bottom cylinder among all simulations of open-loop control as well. The average $C_{D1}$ are -12.02 and -12.01 for DRL-based control and open-loop control, while the average $C_{L3}$ are 24.92 and 24.82, respectively. The wake deflects to the up direction as the counterclockwise rotation of three cylinders. As discussed in the last section, such a smooth, up-deflecting wake can achieve a significant lift and small drag. The above rotation pattern is similar to the Magnus effect according to Ishar \textit{et al.} \cite{ishar2019metric}.

The DRL-based control can achieve the same performance and solution as the open-loop control method in the force extremum problem. It is interesting to explore if $C_{D0}$ on the system can be controlled to track different expected values dynamically. We have tested four cases to demonstrate the better ability of the DRL-based flow control algorithm for drag tracking problems on different $C_{D0}^{ref}$ than the open-loop control method. For DRL control, it is essential to approximately design the reward function $r$ to calculate reward values when the agent inquires about the states. The results of the $C_{D0}$, actions, and vortical wakes are plotted in FIG. \ref{track}, while the setup of the reward in these cases is as equation (3):

\begin{equation}
    r=-(C_{D0}-C_{D0}^{ref})^{2},
\end{equation}

where $C_{D0}$ is the time varying values at each step and $C_{D0}^{ref}$ is the objective of tracking. To test the ability to drag tracking for arbitrary values, we made $C_{D0}^{ref}\in [0,1,2,3]$, respectively. The following results demonstrated that $C_{D0}$ always achieves the $C_{D0}^{ref}$ fluctuating in small ranges in steady after a period of adjustment.

\begin{figure}
\centerline{\includegraphics[width=0.5\textwidth]{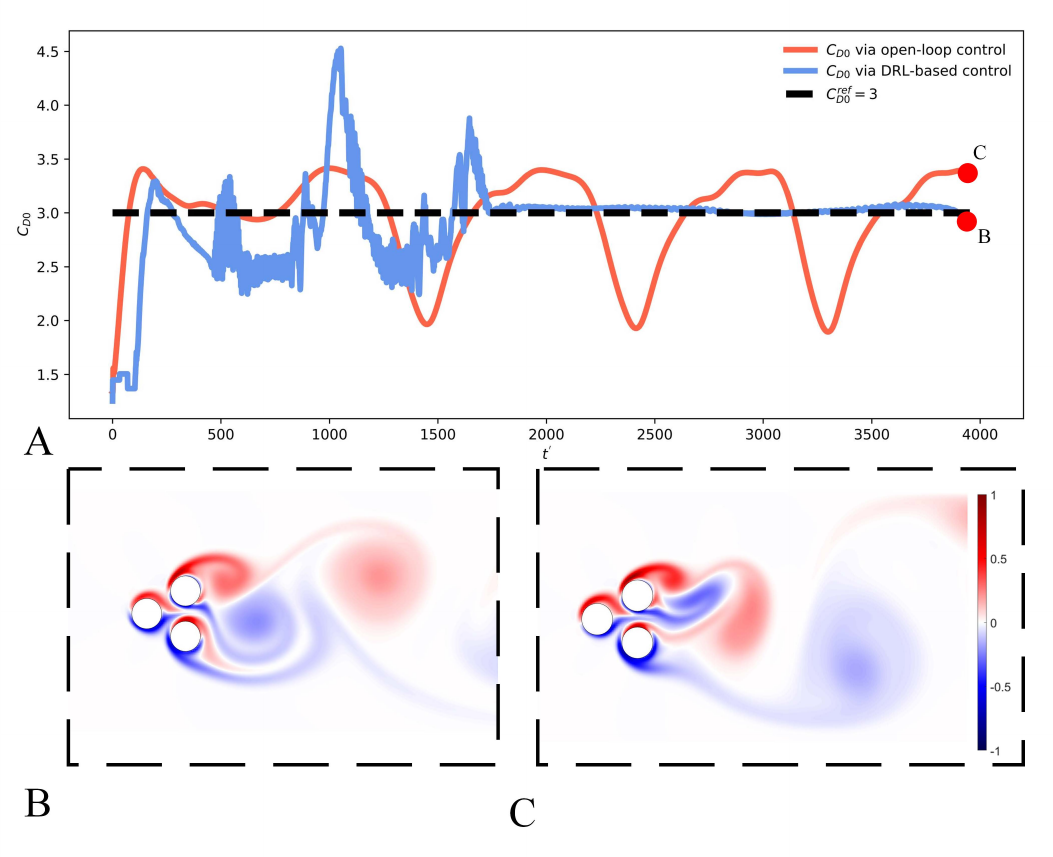}}
\caption{\textbf{Comparison for tracking of $C_{D0}$ keeping $C_{D0}^{ref}=3$.} (A) The curve of $C_{D0}$ changes via DRL-based control (orange line) and open-loop control (blue line). (B) and (C) are the vorticities in the last time step via DRL-based control and open-loop control, respectively sharing the same color bar, which shows DRL can achieve the better policy to retain $C_{D0}$, and its wake is generally different but with many similarities compared with the open-loop method.}
\label{compare}
\end{figure}

In FIG. 7A, the curve of $C_{D0}$ (red line) in the time-domain starts from the initial value to the terminal value fluctuating near 3 with an error less than $\pm0.15$ after 2000 steps in steady. The average $r$ calculated by equation (3) converges to -0.0014, approximately 0. The curve of $C_{D0}$ (green line) in the time domain starts from the initial value to the terminal value fluctuating near 2 with an error less than $\pm0.05$ after 500 steps in steady. The average $r$ converges to -0.0009 in the test experiment. The orange curve when $C_{D0}^{ref}=1$ successfully remains the objective fluctuating about 1 with an error less than $\pm0.05$. The average $r$ converges to -0.0019 in the test. Moreover, the blue line when $C_{D0}^{ref}=0$ also remains the objective fluctuating near 0 with an error less than $\pm0.2$, and the average $r$ during the test is -0.0108.

In FIG. 7B, 7C, 7D, and 7E, the vortical wakes are plotted to illustrate their patterns at the final time, and the rotations of three cylinders have periodic fluctuations similar to that of $C_{D0}$. We can observe that different tracking objectives have different wakes and the DRL algorithm tries to control wakes to unknown patterns with changing action dynamically so that $C_{D0}$ can achieve the expected objectives persistently.

\subsection{\label{sec:level2}Further Remarks on Learning Results}

As mentioned earlier, the results of open-loop control with constant rotation speed are considered the reference baselines of the novel DRL-based feedback control. After analyzing the $C_{D0}$ curves, the rotations' patterns, and the changes of vortical wakes generated by the DRL algorithm, we compared the performances of the above two control methods.

\begin{figure*}
  \centerline{\includegraphics[width=1\textwidth]{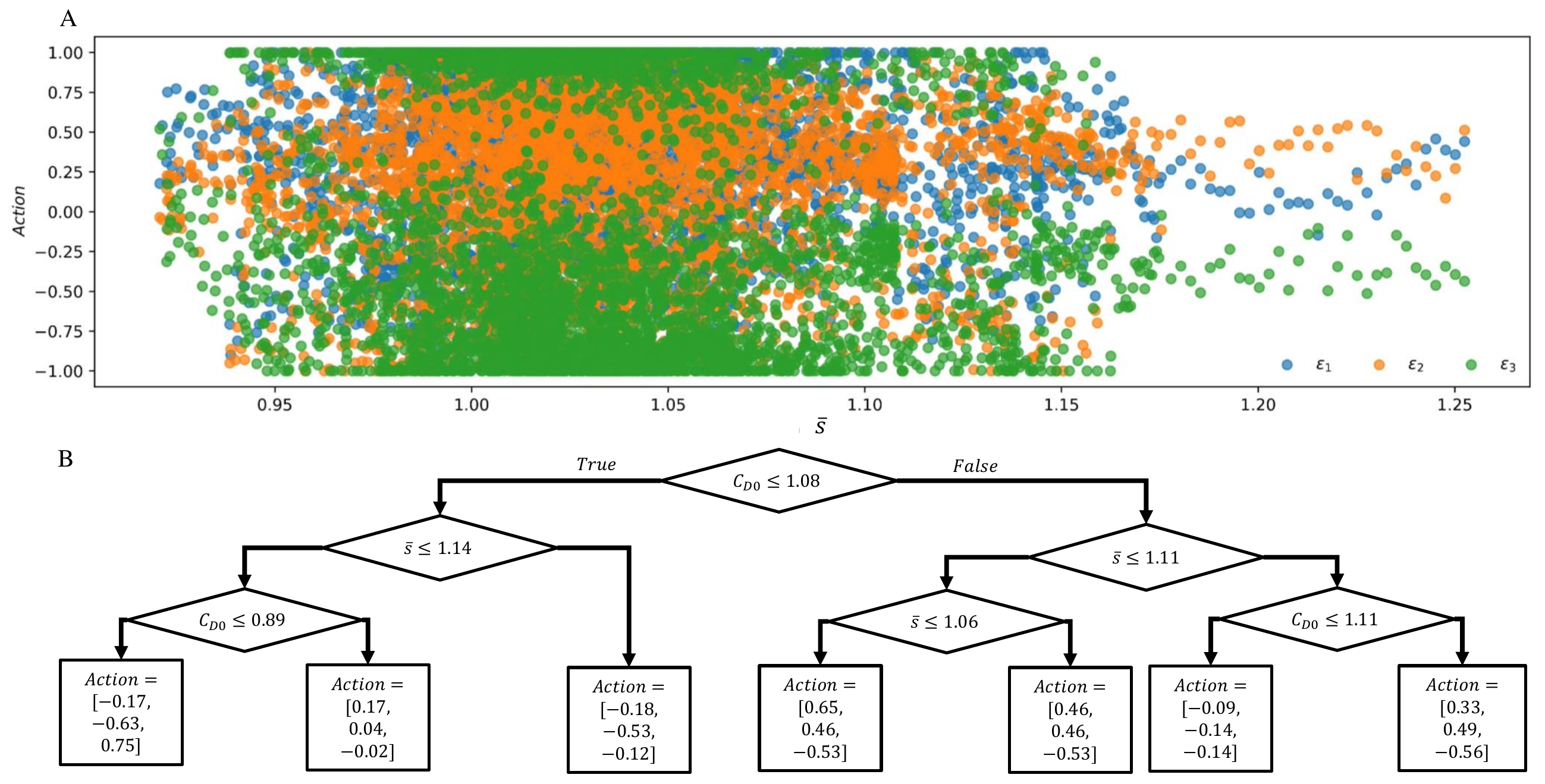}}
  \caption{\textbf{Actions' distribution, and its correspondence with $C_{D0}$ and $\overline{s}$} (A) The distribution of $\varepsilon_{1}$, $\varepsilon_{2}$, $\varepsilon_{3}$ along with changes of $\overline{s}$. (B) The decision tree model trained via ($C_{D0}$, $\overline{s}$, $\varepsilon_{1}$, $\varepsilon_{2}$, $\varepsilon_{3}$) by using the package \textit{tree.DecisionTreeRegressor} in \textit{sklearn} where depth is 3 without random state. The proportion of training and test samples is $8:2$, and the tested $MSE$ score is 0.26.}
\label{explain}
\end{figure*}

We analyzed the $C_{D0}$ tracking performance of two kinds of methods to verify that the DRL-based control can retain an accurate force with less heave, which shows the accurate control ability of the DRL algorithm for the fluid dynamic setting. We selected the actions in FIG. 7E and calculated the average actions for each cylinder as the inputs of the open-loop control. The actions are (0.0, -0.3, 0.6). The changes of $C_{D0}$ from the initial state to the expected value based on two control methods is shown in FIG. 8A. The $r$ calculated via equation (3) are -0.0014 and -0.2373 for DRL-based control and open-loop control, while the $\tilde{C}_{D0}$ are 0.029 and 0.492 respectively, after steady. The system driven by open-loop control cannot achieve the $C_{D0}^{ref}=3$, which fluctuates near 2.6 and has a gap to the expected $C_{D0}$, while the result of DRL-based control converges to the objective more accurately. From the wakes generated by two methods in FIG. 8B and 8C, the pattern of wake and vortex shedding by the DRL-based control method is generally different but with many similarities compared with the open-loop method. Thus, the DRL-based control learns the novel policy to track the force reference more precisely with the real-time feedback tune, which is reasonable to achieve better performance. It is also the advantage and importance of DRL-based real-time feedback flow control, as it is able to respond to various instant possible changing transitions of \textit{fluidic pinball} even the small changes.

Although the DRL-based control can achieve better performance, it is still being determined how DRL found the optimal solution and generates such reasonable actions. Thus, we attempted to enlighten why actions are generated reasonably by DRL-based control with the help of open-loop control via brute-force searching results and fluid dynamics knowledge. In the problem which aims to minimize the $C_{D1}$ and maximize the $C_{L3}$, it was easier to shed light on the decision-making after convergence. After about 800 $t^{'}$, all actions became -1, and three cylinders rapidly rotated counterclockwise. It means the agent has already learned the law of such rotations and can achieve the minimum $C_{D1}$ and maximum $C_{L3}$ by thousands of simulations, in which the patterns of rotations and wake are the same as the result of open-loop control. Then, no matter what feedback $C_{D1}$ and $C_{L3}$ were, rotations did not change and always kept the values. However, for the more complicated force tracking problems, it is much more difficult to comprehend the behavior of decision-making as actions are time-vary and dynamic. We introduced a decision tree model to comprehend the basis of actions made by DRL.

Decision tree (DT) is a type of machine learning model to realize the prediction function with supervised learning, which has often been used to create easy-to-understand solutions to classification and regression problems \cite{loh2011classification, alsagheer2017popular}. It is a tree that starts at a root node and branches based on conditions. An additional benefit to DT is that they can be represented graphically, which aids in human understanding. Some work has explored to use DT model to show the interpretability of the deep neural network (DNN) and DRL \cite{gunning2019xai, mahbooba2021explainable, bastani2018verifiable, puiutta2020explainable, vouros2022explainable}. In this work, we applied a DT model to illuminate which decisions were taken in which situations using the data from a well-trained agent for the $C_{D0}^{ref}=1$ tracking problem. In the beginning, the agent was employed to collect the data ($C_{D0}$, $\overline{s}$, $\varepsilon_{1}$, $\varepsilon_{2}$, $\varepsilon_{3}$), where $\overline{s}$ is the average value among each state with current feedback and historical $C_{D0}$. Then, we used the data set to train the DT model with $C_{D0}$ and $\overline{s}$ as input and $\varepsilon_{1}$, $\varepsilon_{2}$, $\varepsilon_{3}$ as labels. After training, we can apply the trained DT model to imitate actions decided by input situations, and the mean square error ($MSE$) was about 0.26, which means the trained DT model fitted the mapping of data well shown in FIG. \ref{explain}. 

From FIG. 9A, the distribution between $\varepsilon_{1}$, $\varepsilon_{2}$, $\varepsilon_{3}$ and $\overline{s}$ formed the clear dividing that $\varepsilon_{2}>0$ and $\varepsilon_{3}<0$ in the right side when $\overline{s}>1.15$, which made shears around cylinders disperse outside, leading to the smooth and narrow wake reduce $C_{D0}$ (boat-tailing mechanism). However, we cannot observe a clear action trend only via a $\overline{s}$ when $\overline{s}$ is small. By using the DT model further integrating the $C_{D0}$ of each step shown in FIG. 9B, the root node divides the branch into two parts according to whether the $C_{D0}$ is greater than the $C_{D0}^{ref}=1.08$. When $C_{D0}>1.11$ and $\overline{s}>1.11$, that means $C_{D0}$ and $\overline{s}$ are larger than the expected value; the action (0.33, 0.49, -0.56) was taken similarly as the phenomenon in 9A. The base-bleed mechanism (-0.17, -0.63, 0.75) leading to the complex and wide wake happened when $C_{D0}<0.89$, although $\overline{s}$ did not show some characteristics. Due to the complexity of DRL, there are thousands of probabilities to generate action. The purpose of our method was not to reveal the one-to-one correspondence between each state and action, but to apply such sample DT model to shed light on the basis of decision-making obeying human understanding. 




\section{\label{sec:level1}Conclusion}

We illustrated that the forces of both drag and lift on the system composed of three bluff cylinders (\textit{fluidic pinball}) could be controlled to various objectives by applying DRL-based feedback flow control. The well-trained DRL-based control method finds the optimal policy to achieve the minimum drag and maximum lift, the same as the brute-force searching, and explores the novel control policy and wake pattern to track any expected drags. Furthermore, we systematically analyzed the force distribution and vortical wake pattern of the \textit{fluidic pinball}, which provides the detailed baseline and reference to measure the performance of control approaches. Moreover, we explored to understand the basis of decision-making in DRL-based control for fluid dynamic problems, and the treelike-structure machine learning model can shed light on the decisions generation according to the hydrodynamic forces feedback. 

The applications of DRL in fluid dynamics are encouraging, while there are also some gaps waiting for exploring and solving. In this work, only the force information was extracted as a state to learn, and we are studying the probability of higher dimensional force and wake control approach based on DRL for \textit{fluidic pinball}, applying the detailed pressure and velocity information from the wake. On the other hand, we have developed a distributed computing framework based on this work, which can solve CFD problems in a significant number of fluid environments at the same time. We are embedding such distributed computing framework in DRL-based flow control. Currently, we are also exploiting more efficient and detailed method to reveal decision-making for DRL based on fluid dynamic problems, and find out better regulation to design the state and reward function.

\begin{acknowledgments}
We would like to acknowledge the funding supported by State Kay Laboratory of Ocean Engineering (Shanghai Jiao Tong University) (Grant No. GKZD010081) and the research initiation grant provided by the Westlake University (No. 103110556022101). 
\end{acknowledgments}

\section*{AUTHOR DECLARATIONS}
\subsection*{Conflict of Interest}
The authors have no conflicts to disclose.

\subsection*{Author Contributions}
\textbf{Haodong Feng}: Data curation; Visualization; Investigation; Writing – original draft. \textbf{Yue Wang}: Conceptualization (equal); Writing – review and editing. \textbf{Hui Xiang}: Visualization; Writing – review and editing. \textbf{Zhiyang Jin}: Funding acquisition; Writing – review and editing. \textbf{Dixia Fan}: Conceptualization; Funding acquisition; Writing – review and editing.


\section*{DATA AVAILABILITY}
The data and code that support the findings of this study are openly available in Github at \url{https://github.com/HDFengChina/How-to-Control-Hydrodynamic-Force}.

\appendix

\section{Validation of Numerical Method}

\begin{figure}
  \centerline{\includegraphics[width=0.5\textwidth]{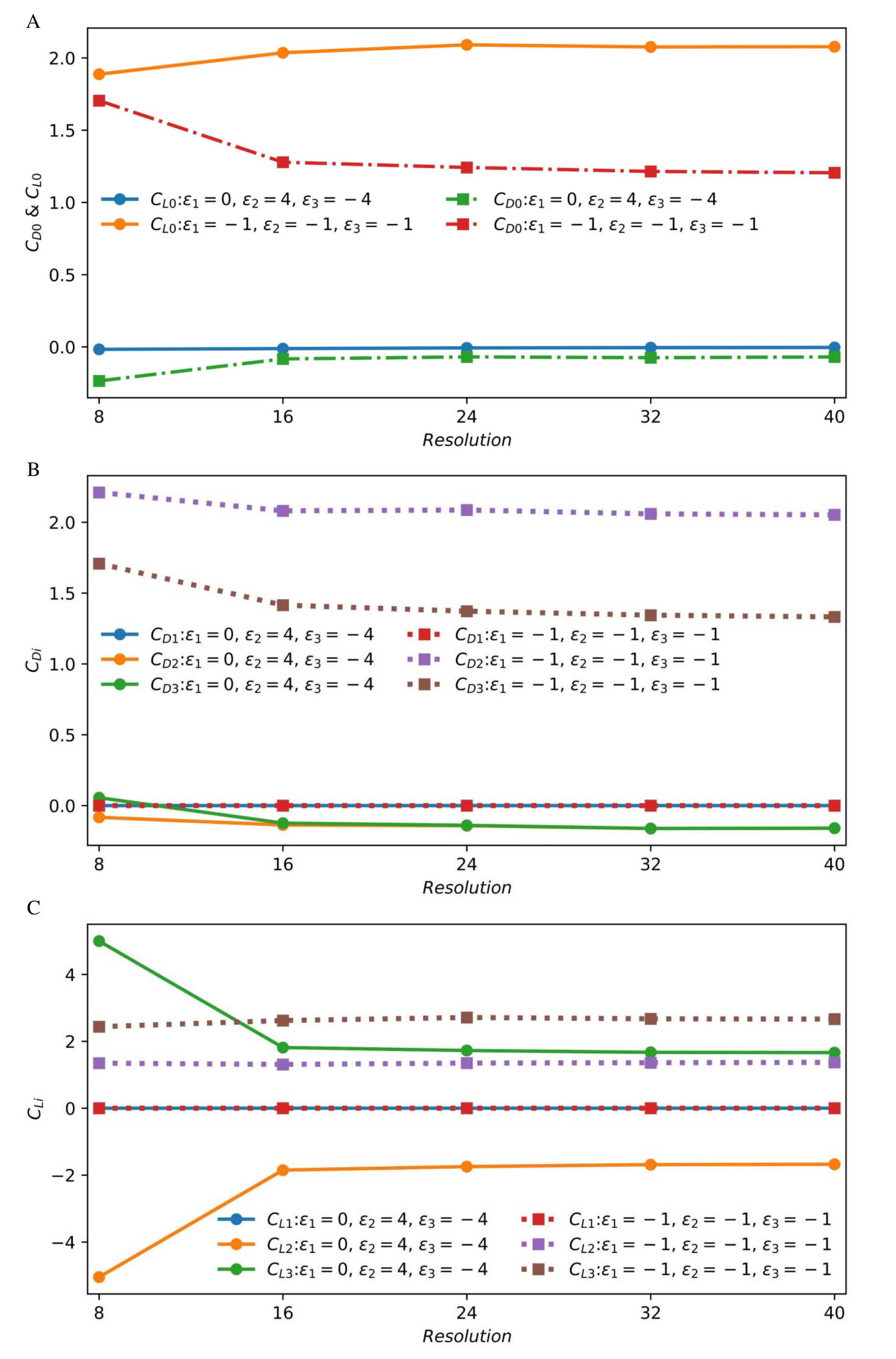}}
  \caption{\textbf{Test the effect of resolution on $C_{Di}$ and $C_{Li}$ in CFD.} (A) The $C_{D0}$ and $C_{L0}$ effected by various resolutions. (B) and (C) The $C_{Di}$ and $C_{Li}$, $i=1,2,3$ respectively.}
\label{curve appendix}
\end{figure}

Different resolutions in the CFD solver are tested to trade between the accuracy and calculation time. We made resolution equal to 8, 16, 24, 36, and 40 to calculate $C_{Di}$ and $C_{Li}$ for cases of boat tailing and the Magnus effect, respectively. The calculation time is significantly increasing along with the increase in resolution. We can observe that $C_{Di}$ and $C_{Li}$ keep similar values when $resolution\geq 16$ from FIG. \ref{curve appendix} while the values are discrepant when $resolution=8$. Thus, we selected $resolution=16$ in this work to speed up the training process with also the relatively accurate calculation.

\begin{figure*}
  \centerline{\includegraphics[width=1\textwidth]{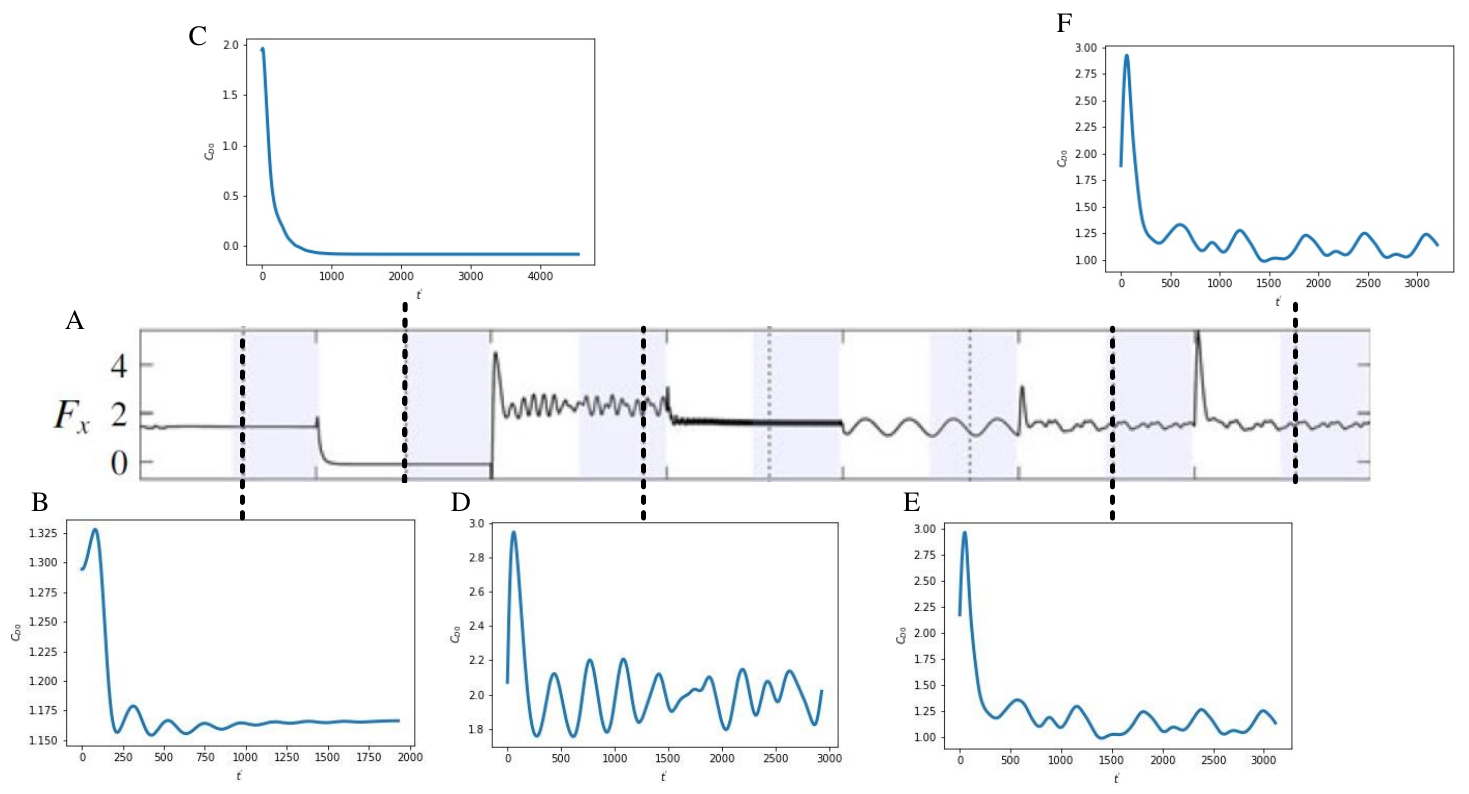}}
  \caption{\textbf{Verification of the CFD environment with previous work.} (A) The screenshot from \cite{ishar2019metric} as the reference of our CFD environment. (B), (C), (D), (E), (F) The $C_{D0}$ of the system corresponds to the unforced no rotation, boat tailing, base bleed, and Magnus effect.}
\label{com appendix}
\end{figure*}

The numerical method applied in this work is feasible to represent the performance of $C_{Di}$ and $C_{Li}$ when $resolution=16$. We verified it with previously published paper \cite{ishar2019metric} as a reference in some different control mechanisms, including the boat tailing, base bleed, Magnus effect, and unforced no rotation shown in FIG. \ref{com appendix}. By comparing the $C_{D0}$ with corresponding references, the results from our solver can accurately represent the trend of these values. The purpose of this work is not to achieve high-accuracy values but to show the control ability of DRL in the CFD environment.

\section{Hyper-parameters in the DRL-based flow control}

The detailed hyper-parameters used by DRL-based flow control in our work are listed in TABLE I.

\begin{table*}[!t]
\begin{center}
\caption{List of hyper-parameters used in DRL}
\begin{tabular}{l r|l r}

\hline
Hidden Layers  & 2 (actor \& critic) & State Dimension & 32 (tracking)/64 (extremum) \\
Exploration Noise  & $\mathcal N(0,0.1)$ & Number of Neurons in Hidden Layers & 256 (actor \& critic) \\
Target Update Rate & $5\cdot10^{-3}$ & Critic Noise & $\mathcal N(0,0.2)$ \\
Buffer Size  & $5\cdot10^{5}$ & Optimizer & Adam \\
Actor Learning Rate  & $10^{-4}$ &  Critic Learning Rate & $10^{-4}$\\
Action Dimension  & 3 & Discount Factor & 0.99  \\

 Number of Environment  & 1 & Batch Size & 512 \\

Gap of Target Update & 2 & Number of Critic in Target & 2 \\

\hline
\end{tabular}
  
\end{center}
\end{table*}

\section*{REFERENCES}
\bibliography{aipsamp}

\end{document}